\documentclass[letterpaper, 10 pt, conference]{ieeeconf}

\pdfoutput=1
\IEEEoverridecommandlockouts
\usepackage[utf8]{inputenc}
\usepackage[pdftex]{graphicx}
\graphicspath{{./Figures}}
\usepackage{amssymb,amsmath}
\usepackage{epstopdf}
\usepackage{float}
\usepackage[font={small}]{caption}
\usepackage{subcaption}
\usepackage{cite}
\let\labelindent\relax % For incompatibility with ieeeconf
\usepackage{enumitem} 
\usepackage[hang,flushmargin]{footmisc} % For no indentation in thanks footnote
\usepackage{amsthm}
    
    \newtheorem{assumption}{Assumption}
    \newtheorem{definition}{Definition}

    \newtheorem{theorem}{Theorem}
    
    \newtheorem{example}{Example}

\usepackage{xcolor}
\makeatletter
\let\NAT@parse\undefined
\makeatother
\usepackage{hyperref}
\hypersetup{%
    colorlinks=true,
    linkcolor={blue!50!black},
    citecolor={blue!50!black},
    urlcolor={blue!50!black}
}

\def\bmat#1{\left[\begin{array}{#1}}
\def\emat{\end{array}\right]}

\def\R{\mathbb{R}} % Set of real numbers
\newcommand{\N}[2][]{\mathbb{N}_{[#1, #2]}} % Natural numbers from a to b: \N[a]{b}
\def\vv#1{{ \rm \bf{#1}}} % Bold math
\newcommand{\cc}[1]{{\mathcal{#1}}} % Short for mathcal
\newcommand{\T}{^\top} % Transpose
\def\st{{\rm s.t.}} % Text for "subject to"
\def\set#1#2{\left\{  #1 : #2 \right\}}
\def\Sum#1#2{\sum\limits_{#1}^{#2}} % Sum with limits
\def\Sp#1{\mathbb{S}_{\succ}^{#1}} % Set of positive definite matrices of dimension #1
\def\Ssp#1{\mathbb{S}_{\succeq}^{#1}} % Set of positive semi-definite matrices of dimension #1

\def\nx{{n_x}} % Dimension of the state
\def\nu{{n_u}} % Dimension of the input
\def\ny{{n_y}} % Dimension of the output
\def\nw{{n_w}} % Dimension of the disturbances
\def\nc{{n_c}} % Dimension of the constraints
\def\cX{\cc{X}} % State constraints
\def\cZ{\cc{Z}} % System constraints (coupled input-state contraints)
\def\cY{\cc{Y}}
\def\cZs{\cZ_{s}} % Set of [s]trictly reachable steady-states
\def\cYs{\cY_{s}} % Set of [s]trictly reachable setpoints (outputs associated to steady-states)
\def\cZw{\cZ_{\phi_K}} % Tightened constraints (for robust linear MPCT)
\def\cZsw{\cZ_{w}} % Set of [s]trictly reachable steady-states for the robust linear MPCT
\def\cYsw{\cY_{w}} % Set of [s]trictly reachable setpoints for the robust linear MPCT
\def\Vf{V_f} % Terminal cost
\def\Vo{V_o} % Offset cost
\def\cXf{\cc{X}_f} % Terminal set
\def\xr{x_r} % State reference
\def\ur{u_r} % Input reference
\def\yr{y_r} % Output reference
\def\xa{x_a} % Artificial state (steady-state)
\def\ua{u_a} % Artificial input (steady-state)
\def\ya{y_a} % Artificial output (steady-state)
\def\cXt{\cc{X}_a} % Terminal set for tracking
\def\cXtw{\cc{X}_{w}} % Terminal set for tracking (robust formulattion)
\def\elleco{\ell_\text{e}} % Economic stage cost
\def\kf{\kappa_f}%feedback for terminal set

\begin{document}
% Fakesection Title

\title{\LARGE Model predictive control for tracking using artificial references: Fundamentals, recent results and practical implementation}

\author{P.~Krupa$^a$,~J.~Köhler$^b$,~A.~Ferramosca$^c$,~I.~Alvarado$^d$,~M.N.~Zeilinger$^b$,~T.~Alamo$^d$,~D.~Limon$^d$%
    \thanks{$^a$~Computer Science Group, Gran Sasso Science Institute (GSSI), L'Aquila, Italy (e-mail: \texttt{pablo.krupa@gssi.it}).}%
    \thanks{$^b$~Institute for Dynamic Systems and Control, ETH Zürich, Zürich CH-8092, Switzerland (e-mails: \texttt{mzeilinger@ethz.ch},  \texttt{jkoehle@ethz.ch}).}%
    \thanks{$^c$~Department of Management, Information and Production Engineering, University of Bergamo, Via Marconi 5, 24044 Dalmine (BG), Italy (e-mail: \texttt{antonio.ferramosca@unibg.it}).}%
    \thanks{$^d$~Department of Systems Engineering and Automation, Universidad de Sevilla, Seville, Spain (e-mails:\texttt{ialvarado@us.es}, \texttt{talamo@us.es},  \texttt{dlm@us.es}).}%
    \thanks{Daniel Limon, Ignacio Alvarado and Teodoro Alamo aknowledge the support of grant PID2022-141159OB-I00 funded by MICIU/AEI/ 10.13039/501100011033  and by ERDF/EU. 
    Johannes K\"ohler was supported by the Swiss National Science Foundation under NCCR Automation (grant agreement 51NF40 180545).
    Pablo Krupa acknowledges the support of the MUR-PRO3 project on Software Quality and the MUR-PRIN project DREAM (20228FT78M).}
}

\maketitle
\thispagestyle{plain}
\pagestyle{plain}

% Fakesection Abstract
\begin{abstract}
This paper provides a comprehensive tutorial on a family of Model Predictive Control (MPC) formulations, known as \emph{MPC for tracking}, which are characterized by including an artificial reference as part of the decision variables  in the optimization problem.
These formulations have several benefits with respect to the classical MPC formulations, including guaranteed recursive feasibility under online reference changes, as well as asymptotic stability and an increased domain of attraction.
This tutorial paper introduces the concept of using an artificial reference in MPC, presenting the benefits and theoretical guarantees obtained by its use.
We then provide a survey of the main advances and extensions of the original linear \emph{MPC for tracking}, including its non-linear extension.
Additionally, we discuss its application to learning-based MPC, and discuss optimization aspects related to its implementation.
\end{abstract}


\section{Introduction} \label{sec:introduction}

The main benefit of Model Predictive Control (MPC)~\cite{Rawlings_MPC_2017, Camacho_S_2013} is its ability to steer the system to a given reference without violating the constraints while minimizing some objective.
A suitably designed MPC controller guarantees asymptotic stability of the closed-loop system to the desired reference as long as its optimization problem is feasible for the initial state of the system.
This is typically achieved by means of a terminal cost function and a terminal invariant set, which must satisfy certain stabilizing conditions in a neighborhood of the reference.

However, it is not unusual for the desired reference to be changed during the online operation of the system, in order to steer it to a more convenient goal.
One of the limitations of classical MPC is that changing the reference online may lead to an unfeasible MPC problem or to the loss of the stabilizing design guarantees.
Furthermore, due to a lack of deep knowledge of the system and the uncertain nature of the change of reference, it is possible for the user to provide a desired reference that is unfeasible or non-attainable for the MPC controller, leading to the same problem.

This tutorial brings together recently-proposed MPC formulations that have been designed to address these issues.
Recursive feasibility, even in the event of a sudden reference change, is guaranteed thanks to the idea of introducing an artificial reference as a decision variable of the MPC's optimization problem.
This property is guaranteed regardless of the reference provided by the user, even if it violates the system constraints or if it is not a steady state of the system.
Additionally, the resulting predictive controller provides asymptotic stability and (in general) a larger domain of attraction when compared to classical MPC.

The adoption of an artificial reference to deal with constraint satisfaction, even in the case of set-point changes, was introduced in the reference governor techniques, where a optimization-based reference filter is used in cascade with a stabilizing controller, see \cite{garone2017reference} and the references therein. 
An artificial reference was first integrated into a generalized predictive controller in \cite{PradaIWC99}, but without stability guarantees.

In 2008, the original stabilizing linear \emph{MPC for tracking} formulation for tracking piece-wise affine references was introduced  \cite{Limon_A_2008, Ferramosca_A_2009}, and in 2018  it was extended to non-linear MPC~\cite{Limon_TAC_2018}. Based on these seminal papers, the idea of using an artificial reference has been extended to other classical MPC paradigms, such as tracking periodic references~\cite{Pereira_PEMPC_2015, Limon_JPC_2014, Limon_MPCTP_2016, Kohler_NMPC_18, Kohler_AUT_2020, yang2021nonlinear, Kohler_IFAC_2023},
economic MPC~\cite{fagiano2013generalized, muller2013economic, ferramosca2014economic, Limon_JPC_2014, Pereira_PEMPC_2015, kohler2020periodic},
robust control~\cite{Limon_JPC_2010, zeilinger2014real, Nubert_RAL_2020, peschke2023robust, morato2024robust},
stochastic MPC~\cite{Paulson_JPC_2019, DJorge_OCAM_2020},
zone-control~\cite{Ferramosca_APC_2010, Ferramosca_JPC_2012, Abuin_JPC_2024},
output-tracking~\cite{Kohler_TAC_2022},
path following and obstacle avoidance~\cite{Sanchez_JIRS_2021, Santos_AUT_2024},
or distributed/coordinated control~\cite{Ferramosca_AUT_2013, Chanfreut_TAC_2021, aboudonia2021distributed, Kohler_IFAC_2023, rickenbach2024active}.

Other articles have provided additional theoretical guarantees on the performance of \emph{MPC for tracking}~\cite{Kohler_CSL_2023}, extended the non-linear case to the use of a semidefinite cost functions~\cite{berberich2020data_semidefinite, Galuppini_IFAC_2023}, presented a novel way of parameterizing the artificial reference as a harmonic signal~\cite{Krupa_CDC_19, Krupa_TAC_2022, Krupa_arXic_ellipHMPC_2023}, focused on a data-driven/learning approach~\cite{berberich2020data_semidefinite, Berberich_TAC_2022, rickenbach2024active, prajapat2024safe}, or worked on the use of soft-constraints~\cite{Zeilinger_TAC_2014, Gracia_softMPCT_arXiv_2024}.

One of the possible drawbacks of these formulations is the additional complexity of their optimization problems due to the inclusion of the artificial reference as decision variables.
Thus, the role of optimization is fundamental to their success, since their theoretical and practical benefits are only useful in practice if they can compete with the classical MPC formulations.
Over the years, several publications have tackled the problem of solving these formulations~\cite{Kohler_AUT_2020, Krupa_TCST_21, Krupa_HMPC_solver_TAC_22, Gracia_ECC_2024, Gracia_softMPCT_arXiv_2024, kohler2024analysis}, presenting solvers and toolboxes~\cite{Spcies} tailored to them.

The \emph{MPC for tracking} formulation and its extensions have also been used in many (academic) case studies and applications, including
automatic insulin injection for diabetes~\cite{Abuin_JPC_2024},
aerospace rendezvous~\cite{Dong_AA_2022, Rebollo_arXiv_2024},
robotics~\cite{Nubert_RAL_2020, Krupa_ECC_21, rickenbach2024active},
or economic building heat and ventilation~\cite{Borja_EaB_2024}, among others.

This article presents a comprehensive tutorial on the use of artificial references in MPC, starting from the presentation of the original linear \emph{MPC for tracking} formulation~\cite{Limon_A_2008, Ferramosca_A_2009} in Section~\ref{sec:fundamentals}, then presenting its main linear extensions and variations in Section~\ref{sec:extensions}, and its non-linear extensions in Section~\ref{sec:nonlinear}.
Section~\ref{sec:learning} presents the application of \emph{MPC for tracking} to learning-based MPC, and Section~\ref{sec:optimization} discusses optimization aspects related to the implementation of these formulations, including guidelines on how to implement them using existing tools from academia.
Final conclusions and remarks are presented in Section~\ref{sec:conclusions}.

% Fakesection Notation
\noindent\textbf{Notation:} The set of natural numbers is denoted by $\mathbb{N}$, and $\N[a]{b}$ denotes the set of integers from the integers $a$ to $b$, both included.
$I_{n}$ denotes the identity matrix of dimension $n$.
We denote $\| x \|_{Q} \doteq \sqrt{x\T Q x}$.
Given two vectors $x$ and $y$, $x \leq (\geq) \; y$ denotes componentwise inequalities.
For vectors $x_1$ to $x_N$, $(x_{1}, x_{2}, \dots, x_{N})$ denotes the column vector formed by their concatenation.
By $\Sp{n}$ and $\Ssp{n}$ we denote the set of positive definite and positive semi-definite matrices in $\R^{n\times n}$, respectively. 
For a set $\mathbb{X}\subseteq\R^n$, $\mathrm{int}(\mathbb{X})$ denotes the interior and for a given $\sigma \in \R$, $\sigma \mathbb{X}=\{ \sigma x \in \R^n:  x \in \mathbb{X} \}$. 
Given sets $\mathbb{X}, \mathbb{Y} \subseteq \R^n$, their Minkowski sum if defined by $\mathbb{X} \oplus \mathbb{Y} \doteq \set{x + y}{x \in \mathbb{X}, y \in \mathbb{Y}}$ and their Pontryagin set difference by $\mathbb{X} \ominus \mathbb{Y} \doteq \set{x}{x \oplus \mathbb{Y} \subseteq \mathbb{X}}$.
A function $f \colon \R \to \R$ is of class $\mathcal{K}_\infty$ if it is continuous, strictly increasing, $f(0) = 0$ and $f(x) \to +\infty$ as $x \to +\infty$.
We use $v_{(i)}$ to denote the $i$-th element of vector $v$.

\section{Fundamentals of MPC for tracking using an artificial reference: the original formulation} \label{sec:fundamentals}

As an introduction to the idea of using an artificial reference in MPC, let us start by considering the original linear \emph{MPC for tracking} piece-wise affine references~\cite{Limon_A_2008, Ferramosca_A_2009}.
As in classical linear MPC~\cite{Rawlings_MPC_2017, Camacho_S_2013}, we consider a controllable system whose dynamics are described by a linear, discrete-time, time-invariant, state-space model
\begin{subequations} \label{eq:sys:linear}
\begin{align} 
    x(t+1) &= A x(t) + B u(t), \label{eq:sys:linear:x} \\
    y(t) &= C x(t) + D u(t), \label{eq:sys:linear:y}
\end{align}
\end{subequations}
where $x(t) \in \R^\nx$, $u(t) \in \R^\nu$, $y(t) \in \R^\ny$ are the state, input and output at sample time $t$, respectively, $A \in \R^{\nx \times \nx}$, $B \in \R^{\nx \times \nu}$, $C \in \R^{\ny \times \nx}$, $D \in \R^{\ny \times \nu}$.
The control objective of the \emph{MPC for tracking} formulation~\cite{Limon_A_2008} is to steer~\eqref{eq:sys:linear} to a desired reference setpoint $y_r \in \R^\ny$ while satisfying the system constraints
\begin{equation} \label{eq:sys:linear:constraints}
    (x(t), u(t)) \in \cZ, \, \forall t,
\end{equation}
where $\cZ \subseteq \R^{\nx + \nu}$ is a closed convex polyhedron that contains the origin in its interior.
The asymptotic convergence properties of the MPC formulations presented in this article require the following notion of \emph{(strictly) rechable setpoints}.

\begin{definition}[Reachable setpoints] \label{def:reachable:setpoints}
The sets of (strictly) reachable steady states and setpoint of system~\eqref{eq:sys:linear} constrained by~\eqref{eq:sys:linear:constraints}, for a given $\sigma \in [0, 1)$, are given by
\begin{subequations} \label{eq:strictly}
\begin{align} 
    \cZs &\doteq \set{(x, u) \in \sigma \cZ}{x = A x + B u} \subseteq \mathrm{int}(\cZ),  \label{eq:strictly:ss} \\
    \cYs &\doteq \set{C x + D u}{(x, u) \in \cZs}. \label{eq:strictly:y}
\end{align}
\end{subequations}
\end{definition}

In classical MPC, the above control objective is achieved by first computing the optimal steady-state $(\xr,\ur) \in \cZs$ satisfying $\yr = C \xr + D \ur$ and then posing the finite-horizon optimal control problem
\begin{subequations} \label{eq:stanMPC}
\begin{align}  
    \min\limits_{\vv{x}, \vv{u}} \;& \Sum{k = 0}{N-1} \ell(x_k, u_k, \xr, \ur) + \Vf(x_N, \xr) \label{eq:stanMPC:cost}\\
    \st & \; x_0 = x(t), \label{eq:stanMPC:initial} \\
        & \; x_{k+1} = A x_k + B u_k, \; k\in\N[0]{N-1}, \label{eq:stanMPC:prediction} \\
        & \; (x_k, u_k) \in \cZ, \; k \in \N[0]{N-1}, \label{eq:stanMPC:cZ} \\
        & \; x_N \in \cXf, \label{eq:stanMPC:terminal}
\end{align}
\end{subequations}
where $\vv{x} \doteq (x_0, x_1, \dots, x_N)$, $\vv{u} \doteq (u_0, u_1, \dots, u_{N-1})$ are the predicted states and control inputs, respectively, along the prediction horizon $N$; the stage cost function is given by
\begin{equation}
\label{eq:quad_stage_cost}
\ell(x_k, u_k, \xr, \ur) \doteq \| x_k - \xr \|^2_Q + \| u_k - \ur \|^2_R,
\end{equation}
with $Q \in \Ssp{\nx}$, $R \in \Sp{\nu}$, the terminal cost is given by $\Vf(x_N, x_r) \doteq \| x_N - \xr \|^2_P$, with $P \in \Sp{\nx}$; and the terminal set $\cXf$ is assumed to contain the desired reference $\xr$.
Let $\vv{x}^* = (x_0^*, \dots, x_N^*)$, $\vv{u}^* = (u_0^*, \dots, u_{N-1}^*)$ be the optimal solution of~\eqref{eq:stanMPC}.
The control law of~\eqref{eq:stanMPC} is $u(t) = u_0^*$.

It is well known that, under a suitable design of the terminal ingredients $\Vf(\cdot)$ and $\cXf$, the MPC formulation~\eqref{eq:stanMPC} steers the system to the desired setpoint if $\yr \in \cYs$ and the initial state $x(0)$ belongs to the feasibility region of~\eqref{eq:stanMPC}~\cite{Rawlings_MPC_2017}.
The above statement highlights two possible drawbacks of~\eqref{eq:stanMPC}:
\begin{enumerate}[label=\roman*)]
    \item If $\yr$ is changed online, then the terminal set $\cXf$ must be recomputed as a suitable admissible invariant set containing the new steady-state $\xr$, which could be computationally expensive.
        Additionally, this could lead to the loss of feasibility of the MPC controller.
    \item The reference $\yr$ must belong to the set of reachable setpoints $\cYs$ (Definition~\ref{def:reachable:setpoints}). In a practical setting, a lack of deep knowledge of the system could result in the selection of a reference $\yr \not\in \cYs$.
\end{enumerate}
To solve these issues, the \emph{MPC for tracking} formulation~\cite{Limon_A_2008} introduces new decision variables $(\xa, \ua) \in \R^\nx \times \R^\nu$ that act as an \emph{artificial reference}, leading to
\begin{subequations} \label{eq:linMPCT} % linear MPCT
\begin{align}  
    \min\limits_{\substack{\vv{x}, \vv{u},\\ \xa, \ua}} &\; \Sum{k = 0}{N-1} \ell(x_k, u_k, \xa, \ua) + \Vf(x_N, \xa) + \Vo(\ya - \yr) \\
    \st & \; x_0 = x(t), \label{eq:linMPCT:initial} \\
        & \; x_{k+1} = A x_k + B u_k, \; k\in\N[0]{N-1}, \label{eq:linMPCT:prediction} \\
        & \; (x_k, u_k) \in \cZ, \; k \in \N[0]{N-1}, \label{eq:linMPCT:cZ} \\
        & \; \ya = C \xa + D \ua, \label{eq:linMPCT:art_ref:yr} \\
        & \; (\xa, \ua) \in \cZs, \label{eq:linMPCT:art_ref:strict} \\
        & \; (x_N, \xa, \ua) \in \cXt, \label{eq:linMPCT:terminal}
\end{align}
\end{subequations}
where $\Vo(\cdot): \R^\ny \to \R$ is the \emph{offset cost function}, $\cXt \subseteq \R^{2\nx + \nu}$ is the terminal \emph{invariant set for tracking} \cite[\S 2.2]{Limon_A_2008}, and~\eqref{eq:linMPCT:art_ref:strict} requires the artificial reference to be a strictly admissible steady state of the system. 
The intuition behind this formulation is to include an artificial reference that acts as a \emph{proxy} of the desired reference.
The formulation steers the system state towards the artificial reference while also penalizing the distance between this artificial reference and the desired one.
The idea is for both the system state and artificial reference to converge towards the desired reference as $t$ increases.

We note that~\eqref{eq:linMPCT:art_ref:strict} considers \emph{strictly} admissible steady states to avoid a possible loss of controllability that could result from the presence of active constraints on the artificial reference~\cite{Limon_A_2008, rao1999steady}.
However, in practice $\sigma$ can be chosen arbitrarily close to $1$, so the artificial reference can be arbitrarily close to the system constraints.

Under a suitable design of its ingredients, as detailed in the following assumption, the \emph{MPC for tracking} formulation~\eqref{eq:linMPCT} provides recursive feasibility and asymptotic stability to the following \emph{optimal reachable reference}.

\begin{definition}[Optimal reachable reference] \label{def:linMPCT:opt_ref}
The \emph{optimal reachable reference} of the MPC for tracking controller~\eqref{eq:linMPCT} for a given $\yr \in \R^\ny$ and $\sigma \in [0, 1)$ is given by
\begin{equation*}
    \ya^\circ = \arg \min_{y \in \cYs} \Vo(y - \yr).
\end{equation*}
\end{definition}

\begin{assumption} \label{ass:linMPCT:design}
Let $Q \in \Ssp{\nx}$ be such that $(Q^{1/2}, A)$ is observable;
$N$ be greater or equal to the controllability index of~\eqref{eq:sys:linear};
$K \in \R^{\nx \times \nu}$ be a state-feedback control gain such that $A + B K$ is Schur;
and $P \in \Sp{\nx}$ satisfy
\begin{equation*}
    (A + B K)\T P (A + B K) - P \preceq -(Q + K\T R K).
\end{equation*}
Let $\Vo(\cdot)$ be a convex, positive-definite, subdifferentiable function with $\Vo(0) = 0$, such that the optimal solution of $\min_{y \in \cYs} \Vo(y - \yr)$ is unique.
Let $\cXt \subseteq \R^{2\nx + \nu}$ be such that, for all $(x, \xa, \ua) \in \cXt$:
\begin{align*}
    (x, K(x - \xa) + \ua) &\in \cZ, \\
    (Ax + B(K(x - \xa) + \ua), \xa, \ua) &\in \cXt.
\end{align*}
\end{assumption}

We note that the invariant set for tracking $\cXt$ can be computed as the (maximal) positive invariant set of a system extended by the artificial reference, as detailed in \cite[\S 2.2]{Limon_A_2008}, that is, following standard procedures for the computation of terminal invariant sets for MPC controllers.
Additionally, a common choice is to take $\Vo$ as a quadratic function $\| \ya - \yr \|^2_S$, with $S \in \Sp{\ny}$, since in this case problem~\eqref{eq:linMPCT} is a quadratic programming (QP) problem.
The following theorems formalize the theoretical properties of~\eqref{eq:linMPCT}.
We refer the reader to \cite{Limon_A_2008, Ferramosca_A_2009} for their proofs.

\begin{theorem}[Recursive feasibility] \label{theo:linMPCT:feasibility}
Let $\hat{\vv{x}} \doteq (\hat{x}_0, \dots, \hat{x}_N)$, $\hat{\vv{u}} \doteq (\hat{u}_0, \dots, \hat{u}_{N-1})$ be any feasible solution of~\eqref{eq:linMPCT} for the current state $x(t)$. Then, problem~\eqref{eq:linMPCT} is feasible for the successor state $A x(t) + B \hat{u}_0$ for any value of $\yr \in \R^{\ny}$.
\end{theorem}

The first major benefit obtained thanks to the use of an artificial reference is that recursive feasibility is guaranteed in the event of a reference change, even if the new reference does not belong to the set of reachable references $\cYs$. Notice that the set of constraints in \eqref{eq:linMPCT} does not depend on the reference $y_r$. 
Additionally, there is no need to recompute any of the ingredients of~\eqref{eq:linMPCT} when the reference is changed.

\begin{theorem}[Asymptotic stability] \label{theo:linMPCT:stability}
Consider~\eqref{eq:sys:linear} subject to~\eqref{eq:sys:linear:constraints} controlled by~\eqref{eq:linMPCT}, and let Assumption~\ref{ass:linMPCT:design} be satisfied.
Assume $x(0)$ belongs to the feasibility region of~\eqref{eq:linMPCT}.
Then, the closed-loop system is stable, fulfills the system constraints, and asymptotically converges to the optimal reachable reference $\ya^\circ$ given by Definition~\ref{def:linMPCT:opt_ref}.
\end{theorem}

The second major benefit is that asymptotic stability to an admissible steady state is guaranteed even if the reference $\yr \not\in \cYs$, i.e., if the reference is unreachable.
In particular, the consequence of Theorem~\ref{theo:linMPCT:stability} is that the closed-loop system will asymptotically converge to $\yr$ if $\yr \in \cYs$.
Otherwise, it will converge to the \emph{optimal reachable reference} $\ya^\circ$ provided by Definition~\ref{def:reachable:setpoints}, i.e., to the strictly reachable setpoint that minimizes its distance to $\yr$ as measured by the offset cost function $\Vo(\cdot)$.
We note that the MPC formulation~\eqref{eq:linMPCT} cannot converge to references with active constraints due to the use of $\sigma < 1$ in $\cYs$, see~\eqref{eq:linMPCT:art_ref:strict}.
However, once again, $\sigma$ can be taken arbitrarily close to $1$, so in practice the closed-loop system can converge to steady states that are arbitrarily close to the system constraints.

Finally, another benefit of~\eqref{eq:linMPCT} is its increased domain of attraction when compared to~\eqref{eq:stanMPC}. This is best seen by considering the following commonly-used particularization of~\eqref{eq:linMPCT}, which considers a terminal equality constraint:
\begin{subequations} \label{eq:equMPCT} % linear MPCT with terminal equality constraint
\begin{align}  
    \min\limits_{\substack{\vv{x}, \vv{u},\\ \xa, \ua}} &\; \Sum{k = 0}{N-1} \ell(x_k, u_k, \xa, \ua) + \| \xa - \xr \|_T^2 + \| \ua - \ur \|_S^2 \\
    \st & \; x_0 = x(t), \label{eq:equMPCT:initial} \\
        & \; x_{k+1} = A x_k + B u_k, \; k\in\N[0]{N-1}, \label{eq:equMPCT:prediction} \\
        & \; (x_k, u_k) \in \cZ, \; k \in \N[0]{N-1}, \label{eq:equMPCT:cZ} \\
        & \; (\xa, \ua) \in \cZs, \label{eq:equMPCT:art_ref:strictly} \\
        & \; x_N = \xa, \label{eq:equMPCT:terminal}
\end{align}
\end{subequations}
where $T \in \Sp{\nx}$ and $S \in \Sp{\nu}$ model a quadratic offset cost $\Vo(\cdot)$.
This formulation avoids the need to compute a terminal invariant set for tracking $\cXt$ and does not require the use of a terminal cost $\Vf(\cdot)$.
Additionally, it uses a quadratic offset cost function, leading to a QP problem with a simple structure that can be exploited by first-order optimization algorithms (see Section~\ref{sec:optimization}).

\begin{figure}[t]
\centering
 \includegraphics[width=\linewidth]{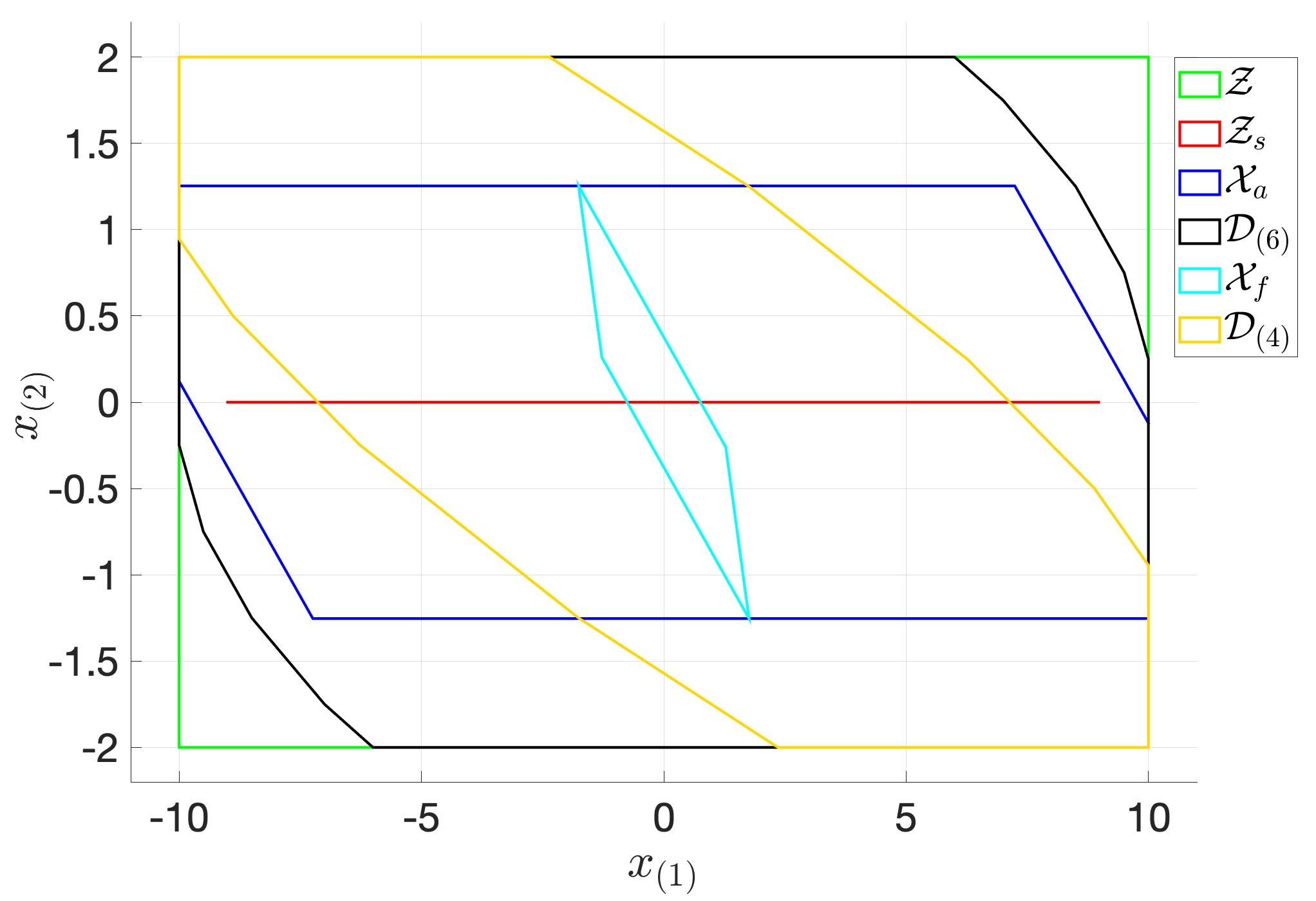}
\caption{Domain of attraction of standard MPC~\eqref{eq:stanMPC} and MPC for tracking~\eqref{eq:linMPCT}. Legend: $\cZ$ are the state constraints; $\cZs$ the states that belong to the manifold of steady states; $\cX_a$ the invariant set for tracking of~\eqref{eq:linMPCT}; $\cX_f$ the terminal invariant set of~\eqref{eq:stanMPC}; $\cc{D}_{(4)}$ and $\cc{D}_{(6)}$ the domains of attraction of~\eqref{eq:stanMPC} and~\eqref{eq:linMPCT}, respectively.}
\label{fig:domain}
\end{figure}

An intuitive way of seeing the increase in the domain of attraction that is typically obtained when using an artificial reference is to compare~\eqref{eq:equMPCT} with its classical MPC counterpart, i.e., with~\eqref{eq:stanMPC} taking~\eqref{eq:stanMPC:terminal} as $x_N = \xr$.
In~\eqref{eq:equMPCT} all states $x(0)$ that can reach \emph{any} setpoint $y \in \cYs$ in $N$ steps belong to its feasibility region, and therefore to its domain of attraction (as per Theorems~\ref{theo:linMPCT:feasibility} and~\ref{theo:linMPCT:stability}).
On the other hand, only states $x(0)$ that can reach the setpoint $\yr$ in $N$ steps may belong to the domain of attraction of the classical MPC counterpart of~\eqref{eq:equMPCT}.
Similar reasoning can be used when comparing the domains of attraction of~\eqref{eq:stanMPC} and~\eqref{eq:linMPCT}, as illustrated by the following example.

\begin{example}[Domain of attraction]
We consider~\eqref{eq:sys:linear} with
\begin{equation*}
    A = \begin{bmatrix}
        1 & 1 \\ 0 & 1
    \end{bmatrix}, \; 
    B = \begin{bmatrix}
        0.5 \\ 1
    \end{bmatrix}, \;
    C = \begin{bmatrix}
        1 & 0
    \end{bmatrix}, \;
    D = \begin{bmatrix}
        0
    \end{bmatrix},
\end{equation*}
and constraints $|x_{(1)}| \leq 10$, $|x_{(2)}| \leq 2$, $|u| \leq 0.5$.
We also consider the classical MPC formulation~\eqref{eq:stanMPC} and the \emph{MPC for tracking} formulation~\eqref{eq:linMPCT}, taking $Q = 100 I_2$, $R = 1$, and $N = 5$.
We take $P$ as the solution of the discrete algebraic Riccati equation and compute the terminal invariant sets of the MPC controllers for the control gain $K$ of the LQR controller.
The terminal set of~\eqref{eq:stanMPC} considers the origin as the reference.
Fig.~\ref{fig:domain} shows the terminal invariant sets and domains of attraction of both controllers.
The results illustrate how the domain of attraction of the \emph{MPC for tracking} formulation can be significantly larger.
This is mostly a result of the difference between both terminal sets.
Indeed, note that the terminal set of~\eqref{eq:stanMPC} is ``centered'' around the reference, whereas the terminal invariant set for tracking of~\eqref{eq:linMPCT} is ``centered'' around the set of feasible steady states $\cZs$, leading to a much larger set.
We note that as the prediction horizon increases, the difference between their domains of attraction becomes smaller.
\end{example}

As a final note, it can be shown that every admissible steady state $(x_s,u_s) \in \mathcal Z_s$ is in the interior of the domain of attraction of~\eqref{eq:linMPCT}.
This means that \emph{MPC for tracking} can stabilize the system for any initial state that is a strictly admissible steady state, which is very interesting from a practical point of view.

\subsubsection*{Local optimality}
A drawback of the inclusion of the artificial reference is that the \emph{MPC for tracking} formulation~\eqref{eq:linMPCT} looses the local optimality property of classical MPC with respect to the unconstrained LQR controller~\cite{hu2002toward}.
That is, the optimal solution of~\eqref{eq:linMPCT} will generally differ from the optimal solution of~\eqref{eq:linMPCT} with the additional constraint $\ya = \yr$ (i.e., its classical MPC counterpart)~\cite{Alvarado_Thesis_07}.
However, this local optimality property can be recovered by a proper selection of the offset cost function $\Vo(\cdot)$.
In particular, local optimality is recovered if $\Vo$ satisfies $\alpha_1 \| y \| \leq \Vo(y) \leq \alpha_2 \| y \|, \, \forall y \in \cYs$, for a sufficiently large $\alpha_1 > 0$.
We refer the reader to \cite[\S 4]{Ferramosca_A_2009} for additional details.

\section{Extensions of MPC for tracking} \label{sec:extensions}

We now present extensions of the original linear \emph{MPC for tracking} formulation~\cite{Limon_A_2008, Ferramosca_A_2009}, to some of the main MPC paradigms: robust MPC, tracking periodic references, and economic MPC.
We focus on presenting formulations that provide a light-weight introduction into the use of artificial references within these MPC paradigms.
Therefore, we focus on linear MPC formulations that resemble the original \emph{MPC for tracking} presented in Section~\ref{sec:fundamentals}.
Section~\ref{sec:nonlinear} will present non-linear extensions of \emph{MPC for tracking}.

\subsection{Robust MPC} \label{sec:extensions:robust}

The use of \emph{MPC for tracking} to robustly control systems with parametric and/or additive uncertainties has been explored using different approaches~\cite{Limon_JPC_2010, zeilinger2014real, Nubert_RAL_2020, peschke2023robust, morato2024robust}.
We recall here the approach presented in~\cite{Limon_JPC_2010}, since it more closely resembles the problem setup presented in Section~\ref{sec:fundamentals}.
This robust \emph{MPC for tracking} formulation is based on the classical tube-based robust MPC for regulation presented in~\cite{mayne2005robust}.

Let us now consider system~\eqref{eq:sys:linear} constrained by~\eqref{eq:sys:linear:constraints} with an additive state disturbance, i.e., we take~\eqref{eq:sys:linear:x} as
\begin{equation} \label{eq:sys:linear:w}
    x(t+1) = A x(t) + B u(t) + w(t),
\end{equation}
where $w(t) \in \R^\nx$ is the unknown state disturbance at time~$t$, which is assumed to be contained within a known compact convex polyhedron $\cc{W}$.
The control objective is to steer the system to a desired reference setpoint $\yr \in \R^\ny$ while satisfying the system constraints~\eqref{eq:sys:linear:constraints} despite the presence of the unknown state disturbance.

The tube-based robust \emph{MPC for tracking} formulation from~\cite{Limon_JPC_2010} is given by
\begin{subequations} \label{eq:RMPCT}
\begin{align}  
    \min\limits_{\substack{\vv{x}, \vv{u},\\ \xa, \ua}} &\; \Sum{k = 0}{N-1} \ell(x_k, u_k, \xa, \ua) + \Vf(x_N, \xa) + \Vo(\ya - \yr) \\
    \st & \; x(t) \in x_0 \oplus \phi_K, \label{eq:RPCT:initial} \\
        & \; x_{k+1} = A x_k + B u_k, \; k\in\N[0]{N-1}, \label{eq:RMPCT:prediction} \\
        & \; (x_k, u_k) \in \cZw, \; k \in \N[0]{N-1}, \label{eq:RMPCT:cZ} \\
        & \; \ya = C \xa + D \ua, \label{eq:RMPCT:art_ref:yr} \\
        & \; (\xa, \ua) \in \cZsw, \label{eq:RMPCT:art_ref:strictly} \\
        & \; (x_N, \xa, \ua) \in \cXtw, \label{eq:RMPCT:terminal}
\end{align}
\end{subequations}
where its ingredients satisfy the following assumption and its control law is given by $u(t) = K(x(t) - x_0^*) + u_0^*$.

\begin{assumption} \label{ass:RMPCT}
The ingredients of~\eqref{eq:RMPCT} satisfy:
\begin{enumerate}[label=(\roman*)]
    \item $Q \in \Ssp{\nx}$, $R \in \Sp{\nu}$, and $(Q^{1/2}, A)$ is observable.
    \item $\Vo(\ya - \yr) = \| \ya - \yr \|_S^2$, with $S \in \Sp{\ny}$.
    \item $K, \bar{K} \in \R^{\nx \times \nu}$ are such that $A_K \doteq A + B K$ and $A_{\bar{K}} \doteq A + B \bar{K}$ are Schur stable matrices.
    \item $\phi_K$ is a robust positive invariant set for the uncertain system $x(t+1) = A_K x(t) + w(t)$, i.e., a set satisfying $A_K \phi_k \oplus \cc{W} \subseteq \phi_k$.
    \item $\cZw \doteq \cZ \ominus (\phi_K \times K \phi_K)$.
    \item $\cZsw \doteq \set{(x, u) \in \sigma  \cZw}{x = A x + B u}$, for some given $\sigma \in [0, 1)$ that is taken arbitrarily close to $1$.
    \item $\Vf(x_N, \xa) = \| x_N - \xa \|_P^2$, with $P \in \Sp{\nx}$ satisfying
    \begin{equation*}
        P - (A + B \bar{K})\T P (A + B \bar{K}) = Q + \bar{K}\T R \bar{K}.
    \end{equation*}
    \item Set $\cXtw \subseteq \R^{2\nx + \nu}$ is an invariant set for tracking for system~\eqref{eq:sys:linear} constrained by $(x(t), u(t)) \in \cZw$ for the control gain $\bar{K}$, i.e., for all $(x, \xa, \ua) \in \cXtw$:
    \begin{align*}
    (x, \bar{K}(x - \xa) + \ua) &\in \cZw, \\
    (Ax + B(\bar{K}(x - \xa) + \ua), \xa, \ua) &\in \cXtw.
    \end{align*}
\end{enumerate}
\end{assumption}

Notice that formulation~\eqref{eq:RMPCT} is similar to~\eqref{eq:linMPCT}.
The main difference if that~\eqref{eq:RMPCT:cZ} considers constraints that are tightened by means of the robust positive invariant set $\phi_K$, whose computation, as well as the computation of a suitable $K$, can be done following standard results~\cite[\S 7]{Limon_JPC_2010}, \cite{mayne2005robust}.
Additionally, the initial predicted state $x_0$ is constrained to a region around $x(t)$, and the terminal invariant set for tracking $\cXtw$ considers the tightened constraints $\cZw$ and a terminal control gain $\bar{K}$ which can be different to the one used for the computation of $\phi_K$.
Indeed, the two gains $K$ and $\bar{K}$ can be designed independently, so as to improve the performance of the controller~\cite[\S 7]{Limon_JPC_2010}.
We refer the reader to~\cite{limon2008design} for guidelines on the design of the ingredients of~\eqref{eq:RMPCT}.

Under the satisfaction of Assumption~\ref{ass:RMPCT}, the robust \emph{MPC for tracking} formulation~\eqref{eq:RMPCT} provides the following recursive feasibility and asymptotic stability guarantees.

\begin{theorem}[\cite{Limon_JPC_2010}, Theorem 1]
Let Assumption~\ref{ass:RMPCT} hold and suppose that~\eqref{eq:RMPCT} is feasible for $x(0)$.
Then, the closed-loop system formed by~\eqref{eq:sys:linear:w}-\eqref{eq:RMPCT} satisfies:
\begin{enumerate}[label=\roman*)]
    \item Recursive feasibility: Problem~\eqref{eq:RMPCT} remains feasible for all $t$, regardless of the realization of $w(t) \in \cc{W}$ and even if $\yr$ is changed online.
    \item Asymptotic stability: If $\yr \in \cYsw$, where
    \begin{equation*}
    \cYsw \doteq \set{C x + D u}{(x, u) \in \cZsw}
    \end{equation*}
    is the set of reachable setpoint considering the tightened constraints $\cZsw$ (cf. Definition~\ref{def:reachable:setpoints}), then the output $y(t)$ converges asymptotically to the set $\yr \oplus (C + D K) \phi_K$.
    Otherwise, output $y(t)$ asymptotically converges to the set $\ya^\circ \oplus (C + D K) \phi_K$, where $\ya^\circ$ is the optimal reachable setpoint considering the tightened constraints $\cZsw$, i.e.,
    \begin{equation*}
    \ya^\circ = \arg \min_{y \in \cYsw} \| y - \yr \|_S^2.
\end{equation*}
\end{enumerate}
\end{theorem}

\subsection{Tracking periodic references} \label{sec:extensions:periodic}

Periodic systems and references naturally occur in many applications, such as water distribution networks~\cite{pereira2016application}, HVAC system~\cite{Borja_EaB_2024}, robotics~\cite{cosner1990plug, romero2022model},  or aerospace~\cite{leomanni2020sum, Dong_AA_2022}.
This has motivated the application of MPC to control periodic systems and/or to track periodic references~\cite{gupta2006period, cao2008repetitive, leomanni2020sum, gondhalekar2011mpc, romero2022model}.
In this subsection, we present the extension of the \emph{MPC for tracking} piece-wise affine reference setpoints presented in Section~\ref{sec:fundamentals} to this control paradigm.
The extension naturally follows by considering a periodic artificial reference.

We focus on the periodic \emph{MPC for tracking} formulation presented in~\cite{Limon_MPCTP_2016}, which can be viewed as an extension of the linear \emph{MPC for tracking} with terminal equality constraint~\eqref{eq:equMPCT} to the problem of tracking a periodic reference.
That is, the control objective is to steer the output $y(t)$ of system~\eqref{eq:sys:linear} to a given periodic reference trajectory $\yr(\cdot) \in \R^\ny$ with known period $\tau \in \mathbb{N}$, i.e., a trajectory satisfying $\yr(t) = \yr(t+\tau)$, $\forall t$, while satisfying the system constraints~\eqref{eq:sys:linear:constraints}.
The notion of reachable setpoints presented in Definition~\ref{def:reachable:setpoints} readily extends to the notion of a reachable periodic reference.

\begin{definition}[Reachable periodic trajectory] \label{def:reachable:periodic}
For a given $\sigma \in [0, 1)$ and period $\tau \in \mathbb{N}$, the set of (strictly) reachable periodic trajectories of~\eqref{eq:sys:linear} constrained by~\eqref{eq:sys:linear:constraints} is
\begin{equation*}
    \cYs^\tau \doteq \set{y(\cdot) \in \R^\ny}{\begin{array}{@{}l@{}} y(t) = C x(t) + D u(t) \\ x(t+1) = A x(t) + B u(t) \\ x(t) = x(t{+}\tau),\, u(t) = u(t{+}\tau) \\ (x(t), u(t)) \in \sigma \cZ \end{array}, \;\forall t}.
\end{equation*}
\end{definition}

The periodic \emph{MPC for tracking} formulation makes use of a periodic artificial reference $(\vv{x_a}, \vv{u_a})$, with period $\tau$, where $\vv{x_a} \doteq (x_{a, 0}, \dots, x_{a, \tau})$, $\vv{u_a} \doteq (u_{a, 0}, \dots, u_{a, \tau - 1})$, leading~to
\begin{subequations} \label{eq:perMPCT} % linear periodic MPCT
\begin{align}  
    \min\limits_{\substack{\vv{x}, \vv{u},\\ \vv{x_a}, \vv{u_a}}} &\; \Sum{k = 0}{N-1} \ell(x_k, u_k, x_{a, k}, u_{a, k}) + V_p(\vv{y_a}, \vv{\yr}) \\
    \st & \; x_0 = x(t), \label{eq:perMPCT:initial} \\
        & \; x_{k+1} = A x_k + B u_k, \; k\in\N[0]{N-1}, \label{eq:perMPCT:prediction} \\
        & \; (x_k, u_k) \in \cZ, \; k \in \N[0]{N-1}, \label{eq:perMPCT:cZ} \\
        & \; x_{a,k+1} = A x_{a, k} + B u_{a, k}, \; k\in\N[0]{\tau-1}, \label{eq:perMPCT:art_ref:dynamics}\\
        & \; (x_{a, k}, u_{a, k}) \in \sigma \cZ, \; k\in\N[0]{\tau-1}, \label{eq:perMPCT:art_ref:constraints} \\
        & \; y_{a, k} = C x_{a, k} + D u_{a, k}, \; k\in\N[0]{\tau-1}, \label{eq:perMPCT:art_ref:yr} \\   
        & \; x_{a, 0} = x_{a, \tau}, \label{eq:perMPCT:art_ref:periodic} \\
        & \; x_N = x_{a, N}, \label{eq:perMPCT:terminal}
\end{align}
\end{subequations}
where $\vv{y_a} \doteq (y_{a, 0}, \dots, y_{a, \tau - 1})$,
\begin{equation*}
    V_p(\vv{y_a}, \vv{\yr}) = \Sum{k = 0}{\tau - 1} \| y_{a, k} - \yr(t+k) \|_S^2, 
\end{equation*}
for $S \in \Sp{\ny}$, is the offset cost function.
Constraints~\eqref{eq:perMPCT:art_ref:dynamics}-\eqref{eq:perMPCT:art_ref:periodic} force the artificial reference $(\vv{x_a}, \vv{u_a})$ to describe a reachable periodic trajectory (Definition~\ref{def:reachable:periodic}) for the given $\sigma \in [0, 1)$, which, once again, is taken arbitrarily close to~$1$.
Finally,~\eqref{eq:perMPCT:terminal} is a terminal equality constraint that forces the predicted terminal state $x_N$ to reach the artificial periodic reference.

The MPC controller~\eqref{eq:perMPCT} shares analogous recursive feasibility and asymptotic stability guarantees to the ones presented in Theorems~\ref{theo:linMPCT:feasibility} and~\ref{theo:linMPCT:stability}, as shown in~\cite{Limon_MPCTP_2016}.
That is, if~\eqref{eq:perMPCT} is feasible for $x(0)$, then the closed-loop system remains feasible at future sample times, even if the reference trajectory $\yr(\cdot)$ is changed to a different periodic trajectory with the same period.
Furthermore, if initially feasible, the closed-loop system converges to the optimal reachable periodic trajectory given by the optimal solution of
\begin{equation*}  
    \min\limits_{\vv{y_a}} \left\{ V_p(\vv{y_a}, \vv{\yr}),\, \st\, \vv{y_a} \in \cYs^\tau \right\}.
\end{equation*}
That is, if $\yr(\cdot)$ describes a reachable periodic trajectory, for the $\sigma$ used in~\eqref{eq:perMPCT}, then the closed-loop system asymptotically converges to it, i.e., $\| y(t) - \yr(t) \| \to 0$ as $t \to +\infty$.
Otherwise, the closed-loop system asymptotically converges to the closest reachable periodic trajectory, as measured by the offset cost function $V_p(\cdot)$.

\subsubsection*{Periodic system}
The periodic \emph{MPC for tracking} formulation~\eqref{eq:perMPCT} can be easily adapted to consider a time-varying periodic system, with period $\tau \in \mathbb{N}$,
\begin{equation} \label{eq:sys:linear:periodic}
    x(t+1) = A(t) x(t) + B(t) u(t),
\end{equation}
where $A(t) = A(t+\tau)$, $B(t) = B(t + \tau)$, $ \forall t$.
In particular, we refer the reader to~\cite{Limon_JPC_2014} for an \emph{MPC for tracking} formulation that considers this paradigm, which naturally arises in several practical applications and has been widely researched within the MPC community~\cite{leomanni2020sum, gondhalekar2011mpc}.

\subsection{Harmonic MPC for tracking} \label{sec:extensions:HMPC}

One of the main disadvantages of~\eqref{eq:perMPCT} is that its number of constraints and decision variables grows with the period $\tau \in \mathbb{N}$ of the reference, since the system dynamics and constraints must be imposed along the complete period of the artificial reference, as shown in~\eqref{eq:perMPCT:art_ref:dynamics}-\eqref{eq:perMPCT:art_ref:constraints}.
This can result in long solve times if the period is too long.
This issue inspired the development of the \emph{MPC for tracking} formulation presented in~\cite{Krupa_CDC_19, Krupa_TAC_2022}, named Harmonic MPC (HMPC).
This formulation was originally developed for tracking piece-wise affine setpoints, although it has recently been extended to the problem of tracking harmonic (aka sinusoidal) reference trajectories~\cite{Krupa_arXic_ellipHMPC_2023}; a particular class of periodic reference that arises in many practical applications.
We now present its original setpoint tracking formulation, since it provides a less notation-heavy introduction of the main idea behind it.

HMPC is an \emph{MPC for tracking} formulation that considers system~\eqref{eq:sys:linear} taking the constraints~\eqref{eq:sys:linear:constraints} as 
\begin{equation*}
    \cZ = \set{(x, u) \in \R^\nx \times \R^\nu}{\underline{y} \leq C x + D u \leq \overline{y}}
\end{equation*}
for some $\underline{y}, \overline{y} \in \R^\ny$ satisfying $\underline{y} < 0 <  \overline{y}$.
As in the \emph{MPC for tracking} formulation~\eqref{eq:equMPCT}, HMPC considers the problem of tracking a given setpoint $(\xr, \ur)$ and makes use of a terminal equality constraint.
However, its artificial reference is given by a harmonic signal, whose value at each prediction step $k$ is given by
\begin{subequations} \label{eq:HMPC:art_ref}
\begin{align} 
    x_{h, k} &= x_{e} + x_{s} \sin(\omega k) + x_{c} \cos(\omega k), \\
    u_{h, k} &= u_{e} + u_{s} \sin(\omega k) + u_{c} \cos(\omega k),
\end{align}
\end{subequations}
where $\omega > 0$ is its frequency and its parameters $x_e, x_s, x_c \in \R^\nx$, $u_e, u_s, u_c \in \R^\nu$ are included as decision variables in the optimization problem.

Using the notation $\vv{x_h} = (x_e, x_c, x_c)$, $\vv{u_h} = (u_e, u_s, u_c)$, $y_e = C x_e + D u_e$, $y_s = C x_s + D u_s$, $y_c = C x_c + D u_c$, the optimization problem of HMPC is given by
\begin{subequations} \label{eq:HMPC} % HMPC
\begin{align}  
    \min\limits_{\substack{\vv{x}, \vv{u},\\ \vv{x_h}, \vv{u_h}}} &\; \Sum{k = 0}{N-1} \ell(x_k, u_k, x_{h,k}, u_{h,k}) + V_h(\vv{x_h}, \vv{u_h}, \xr, \ur) \\
    \st & \; x_0 = x(t), \label{eq:HMPC:initial} \\
        & \; x_{k+1} = A x_k + B u_k, \; k\in\N[0]{N-1}, \label{eq:HMPC:prediction} \\
        & \; \underline{y} \leq C x_k + D u_k \leq \overline{y}, \; k \in \N[0]{N-1}, \label{eq:HMPC:cZ} \\
        & \; x_N = x_e + x_s \sin(\omega N) + x_c \cos(\omega N), \label{eq:HMPC:terminal} \\
        & \; x_e = A x_e + B u_e, \label{eq:HMPC:xe} \\
        & \; x_s \cos(\omega) - x_c \sin(\omega) = A x_s + B u_s, \label{eq:HMPC:xs} \\
        & \; x_s \sin(\omega) + x_c \cos(\omega) = A x_c + B u_c, \label{eq:HMPC:xc} \\
        & \; \sqrt{y_{s(i)}^2 + y_{c(i)}^2} \leq \sigma \overline{y}_{(i)} - y_{e(i)}, \; i \in \N[1]{\ny} \label{eq:HMC:UBy} \\
        & \; \sqrt{y_{s(i)}^2 + y_{c(i)}^2} \leq y_{e(i)} - \sigma \underline{y}_{(i)}, \; i \in \N[1]{\ny}, \label{eq:HMC:LBy} 
\end{align}
\end{subequations}
where $\sigma \in [0, 1)$ is taken arbitrarily close to $1$, the terminal equality constraint~\eqref{eq:HMPC:terminal} enforces the terminal predicted state $x_N$ to reach $x_{h, N}$, i.e., the value of the artificial harmonic reference at prediction time $k = N$, and the offset cost function is given by
\begin{align*}
    V_h(\cdot) = &\| x_e - \xr \|_{T}^2 + \| u_e - \ur \|_{S}^2 \\&+ \| x_s \|_{T_h}^2 + \| x_c \|_{T_h}^2 + \| u_s \|_{S_h}^2 + \| u_c \|_{S_h}^2,
\end{align*}
with $T \in \Sp{\nx}$, $T_h \in \Sp{\nx}$, $S \in \Sp{\nu}$, $S_h \in \Sp{\nu}$, and $T_h, S_h$ diagonal.
In~\cite{Krupa_TAC_2022} it is shown that the equality constraints~\eqref{eq:HMPC:xe}-\eqref{eq:HMPC:xc} impose the satisfaction of the system dynamics on~\eqref{eq:HMPC:art_ref}, whereas the second order cone constraints~\eqref{eq:HMC:UBy}-\eqref{eq:HMC:LBy} impose the satisfaction of the system constraints.
Notice that, even though the HMPC formulation makes use of a periodic artificial reference, the number of constraints of~\eqref{eq:HMPC} does not depend on its period, determined by the choice of $\omega$.

The terms of $V_h$ corresponding to $T$ and $S$ are analogous to the terms from the offset cost function in~\eqref{eq:equMPCT}, in that they penalize a measure of the distance with the desired reference $(\xr, \ur)$.
The other terms of $V_h$ penalize the magnitude of the $\sin$ and $\cos$ terms of the artificial harmonic reference.
The idea is that the ``center'' $(x_e, u_e)$ will converge towards the reference $(\xr, \ur)$, whereas the other parameters of the artificial harmonic reference will converge to $0$.
Indeed, as shown in~\cite{Krupa_TAC_2022}, the HMPC formulation~\eqref{eq:HMPC} shares the same recursive feasibility guarantees and asymptotic stability properties of~\eqref{eq:equMPCT}.
In particular, it is shown that the HMPC formulation asymptotically converges to the optimal reachable reference of~\eqref{eq:equMPCT}, i.e., to the admissible steady state $(x^\circ, u^\circ) \in \sigma \cZ$ that minimizes $\| x^\circ - \xr \|_T^2 + \| u^\circ - \ur \|_S^2$.

The main benefit of the HMPC formulation~\eqref{eq:HMPC}, when compared to~\eqref{eq:equMPCT}, is its increased domain of attraction and performance when working with small prediction horizons, as reported in~\cite{Krupa_CDC_19}.
This benefit is particularly apparent when working with systems that have integrator states and/or slew-rate constraints on the control inputs.

\subsubsection*{Tracking harmonic references}
HMPC~\eqref{eq:HMPC} has been recently extended in~\cite{Krupa_arXic_ellipHMPC_2023} to the problem of tracking a harmonic reference trajectory
of frequency $\omega > 0$, i.e., a reference trajectory whose value at time $t$ is given by
\begin{align*}
    x_r(t) &= x_{re} + x_{rs} \sin(\omega t) + x_{rc} \cos(\omega t), \\
    u_r(t) &= u_{re} + u_{rs} \sin(\omega t) + u_{rc} \cos(\omega t),
\end{align*}
for some parameters $x_{re}, x_{rs}, x_{rc} {\in} \R^\nx$, $u_{re}, u_{rs}, u_{rc} {\in} \R^\nu$.
The formulation is very similar to~\eqref{eq:HMPC}, as well as its properties, which guarantee asymptotic stability to the admissible harmonic reference that minimizes its offset cost function.
Since the complexity of its optimization problem does not depend on the period of the reference, this formulation could be rather useful when dealing with harmonic/sinusoidal references with large periods.

\subsection{Economic MPC for tracking} \label{sec:extensions:economic}

Optimal operation and control of processes is a widely studied field in the overall control community, as well as in the MPC community, where it receives the name of \emph{economic} MPC~\cite{ellis2014tutorial}.
There are several formulations of economic MPC that consider artificial references and/or deal with the problem of changing economic criteria~\cite{fagiano2013generalized, muller2013economic, ferramosca2014economic, Limon_JPC_2014, kohler2020periodic}, as well as several applications, e.g.,~\cite{Pereira_PEMPC_2015, Borja_EaB_2024}.
In this subsection, we recall the economic \emph{MPC for tracking} presented in~\cite{ferramosca2014economic}, since it considers the problem setup from Section~\ref{sec:fundamentals} applied to the problem of minimizing an economic performance metric, given by the function $\elleco(x, u, \theta): \R^\nx \times \R^\nu \times \R^{n_\theta} \to \R$, where $\theta \in \R^{n_\theta}$ are the parameters that characterize the economic criterion.
This economic cost leads to the following notion of the optimal economic setpoint.

\begin{definition}[Optimal economic setpoint] \label{def:economic:setpoint}
The optimal economic setpoint $(x_*, u_*)$ of system~\eqref{eq:sys:linear} subject to~\eqref{eq:sys:linear:constraints} for a given economic cost $\elleco(\cdot)$ parameterized by $\theta$ is given by
\begin{equation} \label{eq:economic:setpoint}
    (x_*, u_*) = \arg \min\limits_{(x, u) \in \cZs} \elleco(x, u, \theta).
\end{equation}
\end{definition}

The economic cost is assumed to satisfy the following standard assumption~\cite{ferramosca2014economic, diehl2010lyapunov}.

\begin{assumption} \label{ass:EMPCT:elleco}
The economic cost $\elleco(\cdot)$ is locally Lipschitz continuous at $(x_*, u_*)$, i.e., there exist scalars $\epsilon, L_e > 0$ such that $\forall \theta$, $\forall (x, u) \in \cZ$, $ \| (x, u) - (x_*, u_*) \| \leq \epsilon$ implies
\begin{equation*}
    \| \elleco(x, u, \theta) - \elleco(x_*, u_*, \theta) \| \leq L_e \| (x, u) - (x_*, u_*) \|.
\end{equation*}
Additionally, for each $\theta \in \R^{n_\theta}$ there exists a multiplier $\lambda$ such that $(x_*, u_*)$ is the unique solution of
\begin{equation*}
    \min\limits_{(x, u) \in \cZ} \ell_r(x, u, \theta).
\end{equation*}
where, additionally, the \emph{rotated} stage cost
\begin{equation*}
    \ell_r(x, u, \theta) \doteq \elleco(x, u, \theta) + \lambda\T(x - (A x + B u)) - \elleco(x_*, u_*, \theta)
\end{equation*}
can be lower bounded by two $\mathcal{K}_\infty$ functions $\alpha_x, \alpha_u$ as
\begin{equation*}
    \ell_r(x, u, \theta) \geq \alpha_x(\|x - x_*\|) + \alpha_u(\|u - u_*\|), \, \forall (x, u) \in \cZ.
\end{equation*}
\end{assumption}

The economic \emph{MPC for tracking} from~\cite{ferramosca2014economic} adds an artificial reference $(\xa, \ua)$ to the classical economic MPC formulation from~\cite{diehl2010lyapunov}, leading to
\begin{subequations} \label{eq:EMPCT} % linear economic MPCT (with terminal equality constraint)
\begin{align}  
    \min\limits_{\substack{\vv{x}, \vv{u},\\ \xa, \ua}} &\; \Sum{k = 0}{N-1} \elleco(x_k {-} \xa {+} x_*, u_k {-} \ua {+} u_*, \theta) {+} \Vo(\xa, \ua, x_*, u_*) \\
    \st & \; x_0 = x(t), \label{eq:EMPCT:initial} \\
        & \; x_{k+1} = A x_k + B u_k, \; k\in\N[0]{N-1}, \label{eq:EMPCT:prediction} \\
        & \; (x_k, u_k) \in \cZ, \; k \in \N[0]{N-1}, \label{eq:EMPCT:cZ} \\
        & \; (\xa, \ua) \in \cZs, \label{eq:EMPCT:art_ref:strict} \\
        & \; x_N = \xa, \label{eq:EMPCT:terminal}
\end{align}
\end{subequations}
where the offset cost $\Vo(\cdot)$ satisfies the following assumption.
Note that~\eqref{eq:EMPCT} requires $(x_*, u_*)$.
Therefore, problem~\eqref{eq:economic:setpoint} must be solved online every time $\theta$ is changed.

\begin{assumption} \label{ass:EMPCT:Vo}
$\Vo(\cdot)$ is a positive definite function such that $(x_*, u_*)$ is the unique minimizer of
\begin{equation*}
    \min\limits_{(x, u) \in \cZs} \Vo(x, u),
\end{equation*}
and there exists a scalar $\gamma > 0$ such that,~$\forall (x, u) \in \cZs$,
\begin{equation*}
    \Vo(x, u, x_*, u_*) - \Vo(x_*, u_*, x_*, u_*) \geq \gamma \| x - x_*\|.
\end{equation*}
\end{assumption}

Formulation~\eqref{eq:EMPCT} provides recursive feasibility guarantees following the same arguments as the original \emph{MPC for tracking} formulation.
Indeed, notice that the constraints of~\eqref{eq:EMPCT} coincide with the constraints of the \emph{MPC for tracking} with terminal equality constraint~\eqref{eq:equMPCT}.
Additionally, the formulation has the following asymptotic stability guarantee.

\begin{theorem}[{\cite[Theorem 1]{ferramosca2014economic}}]
Let Assumptions~\ref{ass:EMPCT:elleco}-\ref{ass:EMPCT:Vo} be satisfied.
Assume that $N$ is greater than the controllability index of system~\eqref{eq:sys:linear}, that there exists a control gain $K_e$ such that $A + B K_e$ has null eigenvalues, and that ${(x_*, u_*) \in \cZs}$.
Then, for a sufficiently large $\gamma$, the closed-loop system formed by~\eqref{eq:sys:linear} and~\eqref{eq:EMPCT} asymptotically converges to $x_*$.
\end{theorem}

We refer the reader to~\cite[Eq. (21)]{ferramosca2014economic} for the lower bound that $\gamma$ must satisfy for the claim in the previous theorem to hold, which depends on the values of $L_e$, $N$ and $K_e$.
We note that, in the case of the economic \emph{MPC for tracking} formulation~\eqref{eq:EMPCT}, asymptotic stability to a reachable steady state is only guaranteed if $(x_*, u_*) \in \cZs$.
Therefore, asymptotic stability is not always guaranteed under online changes of $\theta$, although recursive feasibility is always guaranteed.

\subsubsection*{Periodic economic control}
Economic MPC for tracking has also been extended to periodic control, either due to the use of a periodic economic cost~\cite{Pereira_PEMPC_2015, broomhead2015robust} or to its application to a periodic system~\eqref{eq:sys:linear:periodic}, as in~\cite{Limon_JPC_2014}.

\subsection{Other extensions of MPC for tracking} \label{sec:extensions:other}

As mentioned in the introduction, linear \emph{MPC for tracking} has also been extended to many other control paradigms.
Some notable examples include:
\begin{itemize}
    \item \emph{Stochastic MPC}~\cite{Paulson_JPC_2019, DJorge_OCAM_2020}, which consider an uncertain system and allow a certain level of constraint violation with the use of \emph{chance constraints}~\cite{lorenzen2016constraint}, typically leading to less conservative controllers when compared to robust MPC.
    \item \emph{Zone-control}~\cite{Ferramosca_APC_2010}, where the control objective is to steer the system output to some given set (aka zone), instead of to a specific setpoint. Within the reference set, there is no preference between one point or another. The use of an artificial reference provides recursive feasibility if the reference set is changed online. The extension of zone control \emph{MPC for tracking} to the robust paradigm was presented in~\cite{Ferramosca_JPC_2012}.
    A practical example is the problem of blood-glucose control presented in~\cite{Abuin_JPC_2024}.
    \item \emph{Collision avoidance}~\cite{Santos_AUT_2024}, where obstacles are considered, leading to non-convex constraints.
    \item \emph{Distributed/coordinated control}~\cite{Ferramosca_AUT_2013, Chanfreut_TAC_2021, aboudonia2021distributed, Kohler_IFAC_2023}, where various (coupled) agents that share a common objective are controlled by a distributed optimization algorithm. 
\end{itemize}
 
\section{MPC for tracking with non-linear systems} 
\label{sec:nonlinear}
In this section, we discuss how the \emph{MPC for tracking} formulations naturally generalize to non-linear system dynamics. 
We first explain the setpoint tracking formulation for non-linear systems and discuss the design of terminal ingredients, convexity, and theoretical properties (Sec.~\ref{sec:nonlinear_1}). 
Then, we discuss extensions to periodic reference tracking problems (Sec.~\ref{sec:nonlinear_2}) and robust designs (Sec.~\ref{sec:nonlinear_3}). 
\subsection{Setpoint tracking}
\label{sec:nonlinear_1}
First, we explain the basic non-linear setpoint tracking formulation based on~\cite{Limon_TAC_2018} and then address convexity and the design of terminal ingredients following~\cite{Kohler_AUT_2020,Kohler_TAC_2022}.  
We consider non-linear discrete-time constrained systems 
\begin{subequations} \label{eq:sys:non_linear}
\begin{align} 
\label{eq:sys:non_linear_x}
x(t+1) &= f(x(t),u(t)), \\
\label{eq:sys:non_linear_y}
y(t) &= h(x(t) , u(t)),\\
\label{eq:sys:non_linear_c}
&(x(t), u(t)) \in \cZ, \, \forall t\in\mathbb{N},
\end{align}
\end{subequations}
with $\cZ$ compact and continuous functions $f,h$. 
We define the set of (strictly) feasible steady-state and setpoints
\begin{subequations}
\label{eq:reachable:setpoints_nonlin}
\begin{align}
\cZs&\doteq \set{(\xa,\ua)\in\bar{\cZ}}{\xa = f(\xa , \ua)}     \\
\cYs&\doteq \set{h(\xa,\ua)}{(\xa,\ua)\in\cZs},
\end{align}
\end{subequations}
with some user chosen set $\bar{\cZ}\subseteq \mathrm{int}(\cZ)$. 
Given a reference $\yr$, an optimal setpoint can be computed using
\begin{align}
\label{eq:optimal_setpoint_nonlin}
\min_{\ya\in\cYs}&\Vo(\ya,\yr),
\end{align}
and a minimizer is denoted by $(\xa^\circ,\ua^\circ,\ya^\circ)$. 
Analogous to Problem~\eqref{eq:linMPCT}, the MPC formulation is given by  
\begin{subequations} \label{eq:nonlinMPCT} % non-linear MPCT
\begin{align}  
    \min\limits_{\substack{\vv{x}, \vv{u},\\ \xa, \ua}} &\; \Sum{k = 0}{N-1} \ell(x_k, u_k, \xa, \ua) + \Vf(x_N, \xa,\ua) + \Vo(\ya,\yr) \\
    \st & \; x_0 = x(t), \label{eq:nonlinMPCT:initial} \\
        & \; x_{k+1} = f(x_k ,u_k), \; k\in\N[0]{N-1},  \\
        & \; (x_k, u_k) \in \cZ, \; k \in \N[0]{N-1},  \\
\label{eq:nonlinMPCT_ya}
& \; \ya = h(\xa,\ua),  \\
\label{eq:nonlinMPCT_cxua}
& (\xa, \ua) \in \cZs,\\
        & \; (x_N, \xa,\ua) \in \cXf.
\end{align}
\end{subequations}
We solve Problem~\eqref{eq:nonlinMPCT} at each time $t$ and apply the input $u(t)=u^\star_0$ to system~\eqref{eq:sys:non_linear}. 
For simplicity of exposition, we consider a quadratic cost $\ell$~\eqref{eq:quad_stage_cost} and $\Vo(\ya,\yr)=\|\ya-\yr\|_S^2$, 
with $Q,R,S$ positive definite, see also~\cite{Limon_TAC_2018} for a more general setting. 
Similar to Assumption~\ref{ass:linMPCT:design}, we consider the following conditions on the terminal ingredients $\Vf,\mathcal{X}_{\mathrm{f}}$.
\begin{assumption} \label{ass:nonlinMPCT:design}
There exists a feedback $\kf:\cXf\rightarrow\mathbb{R}^\nu$, such that for any steady-state $(\xa,\ua) \in \cZs$ and any $x\in\mathbb{R}^{\nx}$: $(x,\xa,\ua)\in\cXf$, $u=\kf(x,\xa,\ua)$ satisfies
\begin{itemize}
    \item Constraint satisfaction: $(x,u)\in\cZ$;
    \item Positive invariance: $(f(x,u),\xa,\ua)\in\cXf$;
    \item Control Lyapunov function:\\
      $\Vf(f(x,u),\xa,\ua)-\Vf(x,\xa,\ua)\leq -\ell(x,u,\xa,\ua)$.
\end{itemize} 
\end{assumption}
\begin{theorem}[{Recursive feasibility~\cite[Thm~1]{Limon_TAC_2018}}] \label{theo:nonlinMPCT:feasibility}
Let Assumption~\ref{ass:nonlinMPCT:design} hold and suppose that Problem~\eqref{eq:nonlinMPCT} is feasible at $t=0$. 
Then, Problem~\eqref{eq:nonlinMPCT} is feasible and the closed-loop system satisfies the constraints~\eqref{eq:sys:non_linear_c} for all $t\in\mathbb{N}$, independent of $\yr$.
\end{theorem}
To ensure stability and convergence to the optimal steady-state, we require the following additional conditions.
\begin{assumption}
\label{ass:nonlinMPCT:design_v2}
~\begin{enumerate}[label=\alph*)] 
\item Convexity: The set $\cYs$ is convex.
\label{ass:nonlinMPCT:design_v2_convex}
\item Uniqueness: There exists a unique Lipschitz continuous function $g:\cYs\rightarrow\cZs$, such that for any $\ya\in\cYs$, $(\xa,\ua)=g(\ya)\in\cZs$ satisfy $\ya=h(\xa,\ua)$. 
\label{ass:nonlinMPCT:design_v2_unique}
\item Local upper bound on value function: There exists a constant $\epsilon>0$, such that for any $(\xa,\ua)\in\cZs$ and any $x\in\mathbb{R}^{\nx}$: $\|x-\xa\|\leq \epsilon$, 
Problem~\eqref{eq:nonlinMPCT} is feasible and the optimal cost is quadratically bounded~\cite[Asm.~2]{Kohler_AUT_2020}.
\label{ass:nonlinMPCT:design_v2_local}
\end{enumerate}    
\end{assumption}

\begin{theorem}[{Exponential stability~\cite[Thm.~8]{Kohler_AUT_2020})}] \label{theo:nonlinMPCT:stability}
Let Assumptions~\ref{ass:nonlinMPCT:design}--\ref{ass:nonlinMPCT:design_v2} hold and suppose Problem~\eqref{eq:nonlinMPCT} is feasible at $t=0$. 
Then, the optimal steady-state $\xa^\circ$ is exponentially stable for the resulting closed-loop system. 
\end{theorem}
Next, we discuss the role and constructive satisfaction or relaxation of the conditions in Assumptions~\ref{ass:nonlinMPCT:design}--\ref{ass:nonlinMPCT:design_v2}.  
Note that in the absence of Assumption~\ref{ass:nonlinMPCT:design_v2}, the closed-loop system may in general converge to some steady-state $\xa$, which is not the global minimizer of~\eqref{eq:optimal_setpoint_nonlin}.
\subsubsection*{Design of terminal ingredients}
The simplest design of terminal ingredients uses a terminal equality constraint, i.e., $x_N=\xa$. 
This trivially satisfies Assumption~\ref{ass:nonlinMPCT:design}, while Assumption~\ref{ass:nonlinMPCT:design_v2}\ref{ass:nonlinMPCT:design_v2_local} holds if the linearized dynamics around any steady-state are controllable and the prediction horizon is larger than the controllability index, see~\cite[Sec.~III.A]{Limon_TAC_2018}, \cite[Prop.~4]{Kohler_AUT_2020}. 
In~\cite{Limon_TAC_2018,Kohler_AUT_2020}, a less restrictive terminal cost and set of the form 
\begin{subequations}
\label{eq:terminal_parametrized}
\begin{align}
&\cXf=\set{(x,\xa,\ua)}{\Vf(x,\xa,\ua)\leq \alpha(\xa,\ua)}\\
&\Vf(x,\xa,\ua)=\|x-\xa\|_{P(\xa,\ua)}^2\\
&\kf(x,\xa,\ua)=\ua+K(\xa,\ua)(x-\xa)
\end{align}
\end{subequations}
have been proposed with $P,K,\alpha$ parametrized by the artificial reference.
The standard design of terminal ingredients for non-linear systems involves the linearization at a fixed steady-state, computing $P,K$ based on the LQR, and then choosing a constant $\alpha$~\cite{Rawlings_MPC_2017}. 
To extend this design to MPC for tracking formulations, \cite[App.~B]{Limon_TAC_2018} suggest to partition $\cZs$ and apply a similar design of each partition, resulting in piece-wise constant parametrizations of $P,K,\alpha$. 
This design was improved in~\cite{Kohler_AUT_2020} by deriving a continuous parametrization of $P,K$ using gain-scheduling techniques and determining $\alpha$ implicitly using linear inequality constraints, see also~\eqref{eq:nonlinMPCT_robust} below.

While good terminal ingredients can significantly improve performance, it can also increases the overall complexity of the offline design. One attractive alternative is to remove or relax the requirements on $\cXf,\Vf$. In particular, the terminal set constraint $\cXf$ can be removed if the terminal cost $\Vf$ or the prediction horizon $N$ are chosen sufficiently large~\cite[Sec.~III.B]{Limon_TAC_2018}. 
Furthermore, under a suitable cost-controllability condition, the terminal cost $\Vf$ can be removed if the prediction horizon $N$ is chosen large enough~\cite{Kohler_TAC_2022}. 
This is particularly relevant for applications with non-holonomic vehicles (cf., e.g.,~\cite{rickenbach2024active}), as the linearized dynamics are not stabilizable but an exponential cost controllability condition can still be satisfied by choosing a non-quadratic stage cost, see~\cite{coron2020model} for details. 
\subsubsection*{Convexity and obstacle avoidance}
The (strongly convex) quadratic offset cost $\Vo$ in combination with $\cYs$ convex (cf. Asm.~\ref{ass:nonlinMPCT:design_v2}\ref{ass:nonlinMPCT:design_v2_convex}) is crucial to ensure that the closed-loop system converges to the globally optimal setpoint $\ya^\circ$. 
Notably, convexity is only required in the output space $\cYs$, which is crucial as the steady-state manifold $\cZs$ is often non-convex.  
Nonetheless, especially for motion-planning applications with collision avoidance constraints, $\cYs$ is typically non-convex and hence the system may get stuck at a local minima. 
Different solutions have been proposed to address this problem, such as 
using a homeomorphism to a convex space to define a non-quadratic offset cost $\Vo$~\cite{cotorruelo2020nonlinear} or
choosing a large weighting $\Vo$~\cite{soloperto2023safe}. In~\cite{Santos_AUT_2024}, it was suggested to replace the non-convex constraints with corresponding penalties, which allows for simple implementation with (input-to-state) stability guarantees, see also~\cite[Sec.~5]{Sanchez_JIRS_2021} for a related approach. 
Lastly, in~\cite[Prop.~2]{Kohler_TAC_2022}, a method is presented that ensures convergence to the globally optimal setpoint $\ya^\circ$ for any connected non-convex set $\cYs$. 
This is achieved by defining the offset cost $\Vo$ implicitly as the shortest (feasible) distance using a continuous path $\gamma$:
\begin{align}
\label{eq:distance_non_convex}
&\Vo(\ya,\yr):=\min_\gamma  ~\mathrm{length(\gamma(\cdot))}\\
&\st ~\gamma(0)=\ya,~\gamma(1)=\yr,~\gamma(s)\in\cYs, \forall s\in[0,1].\nonumber
\end{align}
By using a finite parametrization, this cost can be embedded in Problem~\eqref{eq:nonlinMPCT} with additional decision variables and thus ensure convergence for arbitrary (connected) non-convex constraints $\cYs$, see also~\cite[Rk.~3]{Kohler_TAC_2022}.    

\subsubsection*{Uniqueness}
The uniqueness condition (Asm.~\ref{ass:nonlinMPCT:design_v2}\ref{ass:nonlinMPCT:design_v2_unique}) ensures that not only the optimal output $\ya^\circ$ is unique, 
but also the corresponding state and input $(\xa^\circ,\ua^\circ)$. 
This condition can be reduced to a so called trackability condition for the linearized system (cf.~\cite[Rk.~1]{Limon_TAC_2018}).
However, even if this condition is not satisfied, the system will instead converge to the set of optimal steady-states.

\subsection{Tracking periodic references}
\label{sec:nonlinear_2}
In the following, we discuss how the methodology for tracking periodic references (Sec.~\ref{sec:extensions:periodic}) can be extended to non-linear systems based on~\cite{Kohler_AUT_2020}. 
On a conceptual level, this extensions is a relatively straightforward combination of the results for linear periodic tracking (Sec.~\ref{sec:extensions:periodic}) and non-linear setpoint tracking (Sec.~\ref{sec:nonlinear_1}), see~\cite{Kohler_AUT_2020} for corresponding details. 
However, this indirectly comes with additional challenges regarding the design of terminal ingredients and in general an increased online computational demand. 

\subsubsection*{Terminal ingredients}
A method to design terminal ingredients for generic reference trajectories has been proposed in~\cite{Kohler_AUT_2020,kohler2020generic}. 
This approach mirrors the design of continuously parametrized terminal ingredients in~\eqref{eq:terminal_parametrized}, which can be related to a gain-scheduling design for quasi-linear  parameter varying systems. 
Simply choosing a terminal equality constraint is also quite attractive in the non-linear periodic case, as the offline design becomes more challenging. 

\subsubsection*{Computational complexity and partially decoupled tracking-planning}
Another issue inherent to the usage of periodic artificial references is the increased computational demand. 
While this issue is also present for the linear formulation (Sec.~\ref{sec:extensions:periodic}), in the non-linear case this issue is exacerbated as the non-linear optimization problems are computationally more demanding. Furthermore, the Harmonic MPC parametrization (Sec.~\ref{sec:extensions:HMPC}) is difficult to generalize to non-linear system. 
Thus, we present a different methodology based on~\cite[Sec.~3.4]{Kohler_AUT_2020} and \cite[Sec.~4.4]{kohler2024analysis}.  
The main idea is to split the periodic tracking MPC (cf. Problem~\eqref{eq:perMPCT} for the linear case) into two problems:
\begin{itemize}
    \item a periodic reference planner, which optimizes the artificial references $\vv{\xa}, \vv{\ua}$ to track the target $\vv{\yr}$;
\item a trajectory tracking MPC that optimizes $\vv{u}$ to track the artificial reference $\vv{\xa}$. 
\end{itemize}
The main idea is that the computationally intensive long horizon planner does not need to be re-optimized at every time $t$, while the trajectory tracking MPC requires a fast sampling rate to react to disturbances.
The challenge is that running the planner and tracker independently could lead to references that cause feasibility issues in the terminal set constraint of the tracking MPC.
This issue is addressed by adding additional constraints to the planning problem that consider the current tracking error on the terminal set, such that the usual candidate solution remains feasible in the tracker, see~\cite[Sec.~3.4]{Kohler_AUT_2020} for details on a tractable formulation. 
As a result, a standard trajectory tracking MPC can be implemented at each time $t$, while a partially coupled planner can solve a larger optimization problem over a user chosen sampling period. 
The partial coupling ensures that both optimization problems remain recursively feasible. 
Furthermore, if a suitable terminal set is designed, then the reference $\vv{\xa}$ converges to the optimal target in finite time and the state $x$ converges exponentially to the corresponding optimal trajectory $\vv{\xa^\circ}$~\cite[Prop.~13]{Kohler_AUT_2020}. 
This strategy has been further refined in~\cite{benders2024embedded} to address (non-periodic) motion-planning problems in embedded autonomous mobile robots.

\subsection{Robust tracking MPC}
\label{sec:nonlinear_3}
In the following, we focus on designing a robust formulation that ensures safe operation in the presence of bounded model-mismatch. 
This design leverages non-linear robust MPC formulations~\cite{kohler2020computationally,sasfi2023robust,zhao2022tube} which are integrated into the \emph{MPC for tracking} formulation following~\cite{Nubert_RAL_2020}. 
We consider the perturbed dynamics
\begin{align}
\label{eq:sys_nonlin_w}
x(t+1)=f_w(x(t),u(t),w(t)),~w(t)\in\cc{W}\subseteq\R^{\nw}, 
\end{align}
with disturbances $w$ and some known compact set $\cc{W}$ containing the origin. 
The dynamics $f_w$ are assumed to be continuous and we denote the nominal dynamics by $f(x,u)\doteq{f_w(x,u,0)}$. 
Furthermore, we assume that the constraint set is characterized as
\begin{align*}
\cZ=\set{(x,u)}{g_j(x,u)\leq 0,~j\in \N[1]{\nc}}    
\end{align*}
 with Lipschitz continuous scalar functions $g_j$. 
In contrast to linear systems (Sec.~\ref{sec:extensions:robust}), the error propagation for non-linear systems also depends on the nominal dynamics, which makes exact propagation challenging. 
We focus on computationally efficient approaches that provide robust guarantees based on a simple offline computed constraint tightening~\cite{kohler2020computationally,sasfi2023robust,zhao2022tube}. 
To this end, we consider the following conditions on an offline computed stabilizing feedback $\kappa(x,z,v)$ and corresponding (incremental) Lyapunov function $V_\delta(x,z)$. 
\begin{assumption}
\label{ass:increm}
There exist a feedback $\kappa:\R^{\nx}\times \cZ\rightarrow\R^{\nu}$, an incremental Lyapunov function $V_\delta:\R^{\nx}\times\R^{\nx}\rightarrow\R$ and constants $\rho\in[0,1)$, $\bar{w},c_{\mathrm{l}},c_{\mathrm{u}}> 0$, $c_j\geq 0$, $j\in\N[0]{\nc}$, such that for all $(z,v)\in\cZ$, all $w\in\mathcal{W}$ and all $x\in\mathbb{R}^{\nx}$, $u=\kappa(x,z,v)$ satisfies:
%$(x,u)\in\cZs$ with 
\begin{subequations}
\label{eq:increm_lyap}
\begin{align}
\label{eq:increm_lyap_0}
c_{\mathrm{l}}\|x-z\|\leq V_\delta(x,z)\leq&  c_{\mathrm{u}}\|x-z\|\\
\label{eq:increm_lyap_1}
V_\delta(f_w(x,u,w),f(z,v))\leq& \rho V_\delta(x,z)+\bar{w},\\
\label{eq:increm_lyap_2}
g_j(x,u)-g_j(z,v)\leq&  c_j V_\delta(x,z).
\end{align}
Furthermore, the following triangular inequality holds for any $x,z,\xi\in\mathbb{R}^{\nx}$:
%norm-like inequality holds
\begin{align}
\label{eq:increm_lyap_3}
V_\delta(x,\xi)+V_\delta(\xi,z)\geq V_\delta(x,z).    
\end{align}
\end{subequations}
\end{assumption}
Conditions~\eqref{eq:increm_lyap_0}--\eqref{eq:increm_lyap_1} ensures that the feedback $\kappa$ drives the state $x$ exponentially to a neighborhood around any nominal trajectory $z$, which is called incremental stability. 
Inequalities~\eqref{eq:increm_lyap_2} follows naturally if $g_j,\kappa$ are Lipschitz continuous, see~\cite[Prop.~5]{sasfi2023robust} for analytical formulas.
General incremental Lyapunov functions $V_\delta$ and feedbacks $\kappa$ can be constructed offline using (control) contraction metrics~\cite{sasfi2023robust,zhao2022tube}.\footnote{%
Following~\cite[Prop.~2,5]{sasfi2023robust}, condition~\eqref{eq:increm_lyap_1} does in fact not hold for all $x\in\R^{\nx}$ but for all states visited during closed-loop operation due to the tightened constraints.}
In case of only mildly non-linear dynamics, a simple weighted norm $V_\delta(x,z)=\|x-z\|_M$ and linear feedback $\kappa(x,z,v)=v+K(x-z)$ can be constructed, in which case $\bar{w},\rho$ can be computed as (weighted) Lipschitz constants.  
Condition~\eqref{eq:increm_lyap_3} follows naturally if $V_\delta$ is defined as a weighted norm or with a contraction metric. 

Given this offline design, we formulate the following robust \emph{MPC for tracking} formulation from~\cite{Nubert_RAL_2020}:
\begin{subequations}
\label{eq:nonlinMPCT_robust} 
\begin{align}  
    \min\limits_{\substack{\vv{x}, \vv{u},\\ \xa, \ua,\alpha}} &\; \Sum{k = 0}{N-1} \ell(x_k, u_k, \xa, \ua) + \Vf(x_N, \xa,\ua) + \Vo(\ya,\yr) \\
    \st & \; x_0 = x(t), \label{eq:nonlinMPCT_robust:initial} \\
        & \; x_{k+1} = f(x_k ,u_k), \; k\in\N[0]{N-1},  \\
        \label{eq:nonlinMPCT_robust_tighten}
        & \; g_j(x_k, u_k)+c_j\bar{w}\dfrac{1-\rho^k}{1-\rho}\leq 0,\\
        &\;j\in\N[1]{\nc},~ k \in \N[0]{N-1}, \nonumber \\
& \; \ya = h(\xa,\ua),  \\
& \; \xa=f(\xa,\ua),\\
        \label{eq:nonlinMPCT_robust_tighten_steady}
& \;  g_j(\xa, \ua)+c_j\alpha\leq 0,~j\in\N[1]{\nc},\\
        \label{eq:nonlinMPCT_robust_alpha_lower}
& \; \alpha\geq\dfrac{\bar{w}}{1-\rho},\\
        \label{eq:nonlinMPCT_robust_terminal}
& \; V_\delta(x_N,\xa)+\dfrac{1-\rho^N}{1-\rho}\bar{w}\leq \alpha.
\end{align}
\end{subequations}
This formulation initializes the predictions at the measured state $x(t)$ and  the applied control input is $u(t)=u^\star_0$. 
Contrary to the nominal formulation~\eqref{eq:nonlinMPCT}, the constraints~\eqref{eq:nonlinMPCT_robust_tighten} are tightened proportional to the disturbance bound $\bar{w}$. 
The terminal set constraint is characterized by~\eqref{eq:nonlinMPCT_robust_tighten_steady}--\eqref{eq:nonlinMPCT_robust_terminal}, which incorporates an online optimization of the scaling $\alpha$ following~\cite{Kohler_AUT_2020}.
In particular, Condition~\eqref{eq:nonlinMPCT_robust_terminal} defines the terminal set as a sublevel set of the Lyapunov function around the artificial steady-state $\xa$ with an online optimized scaling $\alpha>0$, while 
Condition~\eqref{eq:nonlinMPCT_robust_tighten_steady} ensures that this set lies in the (tightened) constraints. 
Furthermore, Condition~\eqref{eq:nonlinMPCT_robust_alpha_lower} provides a lower bound on the size of the terminal set to ensure robust positive invariance and recursive feasibility. 
Notably, these formulas are particularly simple as the feedback $\kappa$ and Lyapunov function $V_\delta$ from the robust formulation are also used to define the terminal ingredients. 
\begin{theorem}[{\cite[Thm.~1]{Nubert_RAL_2020}}]
Let Assumption~\ref{ass:increm} hold and suppose that Problem~\eqref{eq:nonlinMPCT_robust} is feasible at $t=0$. 
Then, Problem~\eqref{eq:nonlinMPCT_robust} is feasible and the closed-loop system satisfies the constraints~\eqref{eq:sys:non_linear_c} for all $t\in\mathbb{N}$, independent of $\yr$ and for any disturbance realization $w(t)\in\cc{W}$.
\end{theorem}
Given the additional conditions regarding the terminal cost $\Vf$, convexity, and uniqueness (Asm.~\ref{ass:nonlinMPCT:design}--\ref{ass:nonlinMPCT:design_v2}),  and uniform continuity of all the involved functions, this approach also ensures input-to-state stability of the (robustly) optimal steady-state, see~\cite[Thm.~1]{Nubert_RAL_2020} for details.
\subsection{Other extensions and experimental results}
Other notable extensions of the MPC for tracking framework include non-linear MPC formulations that leverage the flexibility of artificial references but aim to directly minimize some economic cost $\elleco$. 
Corresponding results for non-linear systems have been developed in~\cite{fagiano2013generalized, muller2013economic,kohler2020periodic}, see also~\cite[Chap.~5]{kohler2024analysis} for an in depth overview.
Experimental hardware results for non-linear MPC for tracking formulations include 
setpoint tracking with a four-tank~\cite{Limon_TAC_2018}, robotic manipulation with collision avoidance~\cite{Nubert_RAL_2020}, and distributed coverage and collision avoidance with a fleet of miniature race cars~\cite{rickenbach2024active}.
 
\section{Application to Learning-based MPC} \label{sec:learning}
In this section, we highlight how \emph{MPC for tracking} formulations can be a key enabler to address problems in learning-based MPC.  
Learning-based techniques are typically leveraged if the system model or the external environment (objective, constraints) are uncertain, see~\cite{hewing2020learning} for an overview. 
We first provide a general discussion, highlighting where \emph{MPC for tracking} formulations are a crucial component in learning-based approaches. 
Then, we explore two applications in more detail:
\begin{itemize}
    \item safely exploring uncertain constraints~\cite{prajapat2024safe}; 
    \item distributed coverage with unknown objectives~\cite{rickenbach2024active}.
\end{itemize}
\subsubsection*{Learning system dynamics} 
In many applications, the system model~\eqref{eq:sys:non_linear} is subject to significant uncertainty and hence online generated data can be used to improve the model knowledge and thus boost performance.  
However, the fact that our model changes during runtime also poses challenges in determining an optimal steady-state that should be regulated. 
This issue can be naturally addressed with \emph{MPC for tracking} formulations as the desired steady-state is jointly optimized in the MPC. 
Corresponding results for different data-driven and learning-based models can be found in~\cite{Berberich_TAC_2022,strelnikova2024adaptive,peschke2023robust,sasfi2023robust,manzano2020online,wang2022data}. 
 
\subsubsection*{Exploring unknown environment}
Especially in robotics applications, we often do not have an accurate `map' regarding the location of possible obstacles, which are needed to ensure safe exploration with the MPC. 
This problem is typically solved by having a cautious approximation of the obstacle free space which expands by exploring the unknown environment during online operation. 
For such problems, we naturally do not know beforehand where exactly we want to go and in general the targets we may wish to reach can be very far away.  
These problems can be naturally addressed by using \emph{MPC for tracking} formulations, see~\cite{saccani2022multitrajectory,prajapat2024safe,soloperto2023safe}.

\subsubsection*{Guaranteed safe exploration}
In~\cite{prajapat2024safe}, an MPC-based method to systematically guarantee safe exploration of unknown environments is developed. 
In addition to the user-specified system constraints $(x,u)\in\cZ$, safe operation requires $q(y)\leq 0$ where $q:\R^\ny\rightarrow\R$ is an unknown/uncertain function. 
By collecting $n$ (noisy) measurements of this function (e.g., through LIDAR) and given suitable regularity conditions on $q$, one can learn a function $\underline{q}^n(y)\leq q(y)$, which can be used to pessimistically ensure safe operation.  
In particular, this is achieved by leveraging Gaussian processes~\cite{williams2006gaussian}, a versatile machine learning method with guaranteed error bounds. 
During runtime, an MPC determines a location that can be safely reached and where collecting measurements will maximize the information about the uncertain environment.  
This method ensures that (with high probability) the system is safely operated for all times and the unknown domain is explored up to a user chosen tolerance in finite time. 
While the theoretical guarantees largely rely on methods from machine learning and optimal control, the practicality hinges on the utilization of concepts from \emph{MPC for tracking}. 
In particular, persistent safety is ensured by planning trajectories that end in some positively invariant safe set. 
While in principle such a set could be designed offline, this would require a computationally prohibitive prediction horizon to ensure exploration of a large domain. 
These issues are circumvented by relying on techniques from \emph{MPC for tracking} to implicitly define the safe set with an artificial reference. 
In particular, the online learned (pessimistic) constraint $\underline{q}^n(y)$ is used to define a (time-varying) terminal set constraint given by
\begin{align*}
\cXf^n=\{x:~\exists u: (x,u)\in \cZs,~y=h(x,u),~\underline{q}^n(y)\leq 0\}.
\end{align*}
By leveraging techniques similar to Section~\ref{sec:nonlinear}, \cite[Lemma~3]{prajapat2024safe} shows that for path-connected (non-convex) constraints $\cYs$ and locally controllable dynamics, any point in this set can be reached with finite time, which is crucial to ensure exploration of the complete domain in finite time.

\subsubsection*{Distributed coordination and learning}
In~\cite{rickenbach2024active}, a learning-based tracking MPC framework is presented to address the distributed coverage problem. 
The coverage problem requires distributed agents to optimally cover an area, which can be posed as minimizing the coverage cost with partitions $\mathcal{Y}_i\subseteq\R^{\ny}$ and density $\phi:\R^{\ny}\rightarrow\R$:
\begin{align}
\label{eq:learning_coverage}
\min_{y_i,\mathcal{Y}_i}\sum_i\int_{p\in\mathcal{Y}_i} \|p-y_i\|^2 \phi(p) dp.
\end{align}
This formulation can, e.g., be used to model optimal repositioning of a fleet of taxis or aerial robots observing a wild fire.  
The key challenge lies in the fact that methods need to be scalable to a large number of agents, ensure collision avoidance, and deal with uncertain density functions $\phi$.
The framework in~\cite{rickenbach2024active} addresses all these problems simultaneously by leveraging non-linear \emph{MPC for tracking} formulations on multiple levels. 
By adapting the \emph{MPC for tracking} formulation without terminal ingredients from~\cite{Kohler_TAC_2022}, a controller is developed to steer non-holonomic vehicles to arbitrary position references. 
Given these strong properties of the local controllers, distributed coordination is enabled independent of the dynamics of the system by simply coordinating positions and Voronoi partitions based on the Lloyd algorithm. 
Furthermore, a framework is developed to enable efficient exploration and learning. 
This formulation achieves active exploration by using an offset cost $\Vo$ that also depends on the variance of the density function $\phi(y)$ at the artificial steady-state. 

\section{Optimization and aspects and practical implementation of MPC for tracking} \label{sec:optimization}

\subsection{Implementing linear MPC for tracking}

The linear \emph{MPC for tracking} formulations presented in Section~\ref{sec:fundamentals} can be posed as QP problems if the offset cost $\Vo$ is taken as a quadratic or linear function and the invariant set for tracking $\cXt$ is computed as a polyhedron.
From a computational standpoint this is rather advantageous, since QP problems are a particularly simple class of optimization problem for which there are many state-of-the-art efficient solvers, such as~\cite{stellato2020osqp, ferreau2014qpoases, frison2020hpipm}, to name a few.
Furthermore, the resulting QP problem is sparse; a feature that is exploited by most modern solvers~\cite{stellato2020osqp, o2021operator}.
In fact, the resulting sparsity pattern is rather simple, and can thus be exploited by tailored solvers~\cite{Krupa_TCST_21, Gracia_ECC_2024, Spcies}, see Section~\ref{sec:optimization:banded}.

The robust and periodic formulations presented in Sections~\ref{sec:extensions:robust} and~\ref{sec:extensions:periodic}, respectively, are also QP problems under the previous two conditions.
The HMPC formulation presented in Section~\ref{sec:extensions:HMPC} is not a QP problem due to the inclusion of second-order cone constraints~\eqref{eq:HMC:UBy}-\eqref{eq:HMC:LBy}, leading instead to a second-order cone program (SOCP), which can also be rewritten as a quadratically constrained quadratic program (QCQP); both of which can be solved using several state-of-the-art solvers, e.g.,~\cite{o2021operator, frison2022introducing}.
In particular,~\cite{Krupa_HMPC_solver_TAC_22} presents a solver tailored to the HMPC formulation (available in~\cite{Spcies}) and shows that it can be solved in computation times comparable to state-of-the-art QP solvers.

A tool that is rather useful for prototyping and testing linear \emph{MPC for tracking} formulations is the MATLAB toolbox YALMIP~\cite{Lofberg2004}, which provides a simple syntax to build the optimization problem and to then solve it by selecting from a large pool of state-of-the-art solvers, including most of the ones referenced above.

\subsubsection*{Computing the invariant set for tracking}
As mentioned in Section~\ref{sec:fundamentals}, and as discussed in detail in~\cite[\S 2.2]{Limon_A_2008}, $\cXt$ can be obtained as the maximal invariant set of a system obtained by extending~\eqref{eq:sys:linear} with the artificial reference.
This results in a polyhedral set than can be computed using standard procedures~\cite{Rawlings_MPC_2017, fiacchini2010computation}.
The MATLAB MPT3 toolbox~\cite{herceg2013multi} can be rather useful to this end.

Computation of the maximal invariant set can be very expensive.
Several alternatives are typically considered: use a terminal equality constraint, as in~\eqref{eq:equMPCT}; dynamically scale the terminal set~\cite{simon2014reference}; avoid the use of a terminal set by choosing a sufficiently large prediction horizon~\cite{Limon_TAC_2018}; or compute an elliptic terminal invariant set~\cite[\S 4.1]{Blanchini_A_1999} by solving an optimization problem with linear matrix inequality constraints~\cite{wan2003efficient}.
The downside of using an ellipsoid is that the resulting optimization problem is no longer a QP, but instead a SOCP/QCQP.
However, as previously discussed, there are several state-of-the-art solvers well suited for these problems~\cite{o2021operator, frison2022introducing, krupa2024sparse}.

\subsubsection*{Soft constraints}
In spite of the larger domain of attraction typically obtained with the use of artificial references, the presence of constraints can still lead to feasibility issues when dealing with real systems or in the presence of unknown uncertainties.
To avoid feasibility issues, a common practical solution is to make use of soft constraints~\cite{Zeilinger_TAC_2014, Gracia_softMPCT_arXiv_2024}.

\subsection{Exploiting the semi-banded structure of linear MPC for tracking} \label{sec:optimization:banded}
The stage cost $\ell(x_k, u_k, \xa, \ua) \doteq \| x_k - \xa \|^2_Q + \| u_k - \ua \|^2_R$, commonly used in many MPC for tracking formulations, includes cross products between $(x_k,u_k)$ and $(\xa,\ua)$ that break down the strictly banded structure of the optimization problem related to standard MPC \cite{Krupa_TCST_21, krupa2021:PLCs}. Consider, for example, the cost of the  linear \emph{MPC for tracking} formulation with equality constraints~\eqref{eq:equMPCT}:
\begin{equation*}
    V(z) =\Sum{k = 0}{N-1} \ell(x_k, u_k, \xa, \ua) + \| \xa - \xr \|_T^2 + \| \ua - \ur \|_S^2, 
\end{equation*}
where $z$ denotes the vector of decision variables, i.e.,
\begin{equation*}
    z = \left[ x_0\T,\, u_0\T,\, x_1\T,\, u_1\T,\, \ldots,\, x_{N-1}\T,\, u_{N-1}\T,\, \xa\T,\, \ua\T \right]\T.
\end{equation*}
In this case, the cost $V(z)$ is a quadratic function on $z$ that can be rewritten as $V(z)=z\T (H_B+H_{LR
}) z +q\T z+c$, where
\begin{equation*}
    H_B=\bmat{ccccccc} Q & 0 & 
  0 &0&\cdots&0 &0 \\
   0& R & 
  0 &0&\cdots&0 &0 
  \\
  0 & 0 & 
  Q &0&\cdots&0 &0 \\
  0 & 0 & 
  0 &R&\cdots&0 &0 \\
  \vdots& \vdots& \vdots & \vdots & \ddots & \vdots & \vdots\\
  0 & 0 & 
  0 &0&\cdots&NQ &0 \\
     0 &0& 0&0&\cdots&0 &NR\emat,
\end{equation*}
\begin{equation*}
    H_{LR}=\bmat{ccccccc} 0 & 0 & 
  0 &0&\cdots&-Q &0 \\
   0& 0 & 
  0 &0&\cdots&0 &-R 
  \\
  0 & 0 & 
  0 &0&\cdots& -Q &0 \\
  0 & 0 & 
  0 &0&\cdots&0 &-R \\
  \vdots& \vdots& \vdots & \vdots & \ddots & \vdots & \vdots\\
  -Q & 0 & 
  -Q &0&\cdots&T &0 \\
     0 &-R& 0&-R&\cdots&0 &S\emat.
\end{equation*}
We notice that $H_B$ is a banded matrix and $H_{LR}$ is a low-rank matrix.
Thus, the matrix defining the quadratic terms of $V(z)$ has a semi-banded structure that prevents the direct use of the efficient optimization techniques developed for the fully banded case of standard MPC \cite{wang2009fast, krupa2021:PLCs}.
Two different approaches have been presented in the literature to address this issue.
The first approach involves using an extended ADMM formulation that recovers the banded structure by considering three primal decision variables in the ADMM formulation \cite{Krupa_TCST_21}.
The second approach relies on the use of the Woodbury matrix identity~\cite{tylavsky1986generalization}, which allows for solving, in an efficient way, the semi-banded systems of equations required to implement the standard ADMM algorithm (instead of the extended one)  \cite{Gracia_ECC_2024,Gracia_softMPCT_arXiv_2024}.
The obtained computational times are similar to the ones corresponding to standard MPC.

\subsection{Solving non-linear MPC for tracking}
Assuming $f,h$ are smooth, non-linear \emph{MPC for tracking} problems can be efficiently coded with CasADi~\cite{andersson2019casadi} in MATLAB or Python. This provided automatic differentiation and interfaces general purpose sparse non-linear optimization solvers like IPOPT~\cite{wachter2006implementation}. 
For embedded application and long prediction horizons, Acados~\cite{verschueren2022acados} can offer efficient implementation by also leveraging QP solvers that exploit the stage wise structure of MPC~\cite{frison2020hpipm}. 
Problem~\eqref{eq:nonlinMPCT} can be written in this form by augmenting the state space with the artificial reference, as, e.g., done in experiments in~\cite{rickenbach2024active}. 

\section{Conclusion} \label{sec:conclusions}

We have provided a tutorial exposition on \emph{MPC for tracking} formulations, which incorporate an artificial reference as an additional decision variable of the MPC's optimization problem.
The addition of the artificial reference in linear and non-linear MPC provides several important benefits: an increase of the domain of attraction, the ability to deal with non-reachable references, recursive feasibility under online operational changes, and asymptotic stability to the ``best'' reachable objective. 
This paradigm is very versatile, as exemplified by the many extensions developed in the literature, including robust designs, economic costs, periodic references, or zone control.
Furthermore, tailored solvers have been developed that provide solve-times in the order of milliseconds for some of the linear formulations.
Finally, \emph{MPC for tracking} can be a key enabler to address problems beyond tracking in learning-based MPC.

% Fakesection Bibliography
\bibliographystyle{IEEEtran}
\bibliography{IEEEabrv,bib_MPCT_workshop_CDC24}

\end{document}